\RequirePackage{fix-cm}
\documentclass{svjour3}                     
\smartqed  
\usepackage{graphicx}
\usepackage[normalem]{ulem}
\usepackage{mathptmx}      
\journalname{Journal of Computational Neuroscience}

\usepackage{amsmath,amssymb}

\usepackage{changepage}

\usepackage[utf8x]{inputenc}

\usepackage{textcomp,marvosym}

\usepackage[sort]{natbib}

\usepackage{nameref,hyperref}

\usepackage[right]{lineno}

\usepackage{microtype}

\usepackage[table]{xcolor}
\usepackage{color}

\usepackage{array}

\newcolumntype{+}{!{\vrule width 2pt}}

\newlength\savedwidth

\newcommand\thickhline{\noalign{\global\savedwidth\arrayrulewidth\global\arrayrulewidth 2pt}%
\hline
\noalign{\global\arrayrulewidth\savedwidth}}

\renewcommand{\figurename}{Fig}

\usepackage{lastpage,fancyhdr,graphicx}
\usepackage{epstopdf}

\usepackage{amsfonts}
\usepackage{amsmath}
\usepackage{bbm}
\usepackage{dsfont}
\usepackage{bm}

\newcommand{\ind}{{\mathbbm{1}}}
\newcommand{\E}{\mathrm{E}}

\newcommand{\Prob}{\mathrm{Pr}} 
\newcommand{\vect}[1]{\boldsymbol{#1}}

\DeclareMathOperator*{\argmax}{\arg\max}
\DeclareMathOperator*{\argmin}{\arg\min}

\newcommand{\invitro}{\textit{in vitro }}
\newcommand{\invivo}{\textit{in vivo }}
\newcommand{\invivoe}{\textit{in vivo}}

\begin{document}

\title{Monosynaptic inference via finely-timed spikes
}




\author{Jonathan Platkiewicz \and Zach Saccomano \and Sam McKenzie \and Daniel English \and Asohan Amarasingham
}

\institute{Jonathan Platkiewicz* \at
          Department of Mathematics, The City College of New York, The City University of New York, New York, NY 10031, USA \\
           \and
           Zach Saccomano* \at
           Department of Biology,  The Graduate Center, The City University of New York, New York, NY 10016, USA
           \and 
           Sam McKenzie \at
           Neuroscience Institute, New York University, New York, NY 10016, USA
           \and 
           Daniel English \at
           School of Neuroscience, Virginia Tech, Blacksburg, VA 24060, USA
           \and 
           Asohan Amarasingham \at 
           Department of Mathematics, The City College of New York, The City University of New York, New York, NY 10031, USA \\
           Departments of Biology, Computer Science, and Psychology, The Graduate Center, The City University of New York, New York, NY 10016, USA \\
           \email{aamarasingham@ccny.cuny.edu} \\
           {\it *=These authors contributed equally to this work.} \\
           \and
        {\bf~Acknowledgements.} We thank G. Buzs\'{a}ki for providing advice and inspiring our work on this problem, and T. Evans, M. Regnaud, and H. Rotstein for advice and comments. This work was supported by NIMH R01-MH102840 (A.A.), DOD ARO W911NF-15-1-0426 (A.A. and J.P.), PSC-CUNY 68521-00 46 (A.A.), and NIMH K99 MH118423 (S.M.). We warmly acknowledge the hospitality of the Initiative for Theoretical Sciences (ITS) at the CUNY Graduate Center.
}


\maketitle
\begin{abstract}
Observations of finely-timed spike relationships in population recordings have been used to support partial reconstruction of neural microcircuit diagrams. In this approach, fine-timescale components of paired spike train interactions are isolated and subsequently attributed to synaptic parameters. Recent perturbation studies strengthen the case for such an inference, yet the complete set of measurements needed to calibrate statistical models are unavailable. To address this gap, we study features of pairwise spiking in a large-scale {\it in vivo} dataset where presynaptic neurons were explicitly decoupled from network activity by juxtacellular stimulation. We then construct biophysical models of paired spike trains to reproduce the observed phenomenology of \invivo monosynaptic interactions, including both fine-timescale spike-spike correlations and firing irregularity. A key characteristic of these models is that the paired neurons are coupled by rapidly-fluctuating background inputs. We quantify a monosynapse's causal effect by comparing the postsynaptic train with its counterfactual, when the monosynapse is removed. Subsequently, we develop statistical techniques for estimating this causal effect from the pre- and post-synaptic spike trains. A particular focus is the justification and application of a nonparametric separation of timescale principle to implement synaptic inference. Using simulated data generated from the biophysical models, we characterize the regimes in which the estimators accurately identify the monosynaptic effect. A secondary goal is to initiate a critical  exploration of neurostatistical assumptions in terms of biophysical mechanisms, particularly with regards to the challenging but arguably fundamental issue of fast, unobservable nonstationarities in background dynamics. \end{abstract}


\section{Introduction}
\begin{figure}[!htbp]
\includegraphics[width=\linewidth]{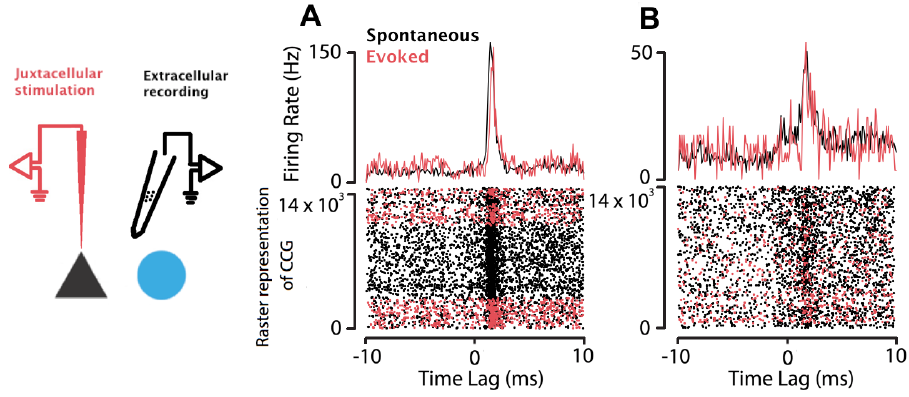}
\caption{{\bf The experiment of \citet{English2017}} \textbf{A:} A putative monosynaptic pair from the data of \citet{English2017} in hippocampal region CA1. {\it Top}: A cross-correlogram (see Methods section \ref{ccg-def}) for a strong monosynaptic connection, normalized by the number of pyramidal spikes. The reference train is the pyramidal
neuron. The red trace is derived from interneuron spikes occurring within -10 to 10 ms of pyramidal spikes
that fire during a juxtacellular stimulation period (‘evoked’ state), while the black trace is derived from interneuron
spikes, in the absence of stimulation (`spontaneous' state). {\it Bottom}: A raster representation of the CCG. Each
row corresponds with the occurrence of a spike of one putatively presynaptic pyramidal neuron. The dots
plotted across a row represent spike times of a (putatively post-synaptic) interneuron relative to the pyramidal
neuron spike, which occurs at lag 0. In this scheme, the same interneuron spike may be represented in
multiple rows. Note the reliability of the effect. Though the `evoked' pyramidal spikes occur during stimulation, there is still an over-abundance of
finely-timed spikes in the interneuron shortly after pyramidal spikes, providing evidence for a monosynaptic
connection. \textbf{B:} The same plots for another putatively monosynaptic pair. In this example, the neuron pair seems to demonstrate a monosynaptic interaction that is riding over a slower co-modulated background signal.}
\label{fig:juxta}
\end{figure}

The difficulty of studying brain connectivity in behavioral conditions is a major challenge in systems neuroscience. One approach is to attempt to infer microcircuit diagrams indirectly from large-scale spike population recordings~\citep{Fujisawa2008}. A synaptic connection is detected when a spike in a reference neuron is followed, with statistical regularity, by finely-timed spikes in a putative target neuron, with (e.g., 2-3 ms) time delays that are consistent with monosynaptic delays. Such a regularity is commonly associated with a millisecond-fine peak in the cross-correlogram (CCG) at lags of 2-3 ms (top panel of Figure \ref{fig:juxta}, black traces). However, there are alternative explanations for these observations, such as indirect polysynaptic partners as well as common (``third party'') input. In this vein, simultaneous optogenetic or juxtacellular stimulation of small groups of cells permit a partial decoupling of presynaptic neurons from the ongoing network activity in \invivo conditions ~\citep{English2017}, and have strengthened the justification for such an analysis (top panel of Figure \ref{fig:juxta}, red traces).

However, even with the benefit of finely-targeted control, the majority of detectable synaptic connections will require analysis of fine-temporal pairwise spike interactions to dissociate coarsely-timed network co-modulation from synaptically-mediated synchrony. Arguably, the state-of-the-art of such inferences is {\it ad hoc} or at least preliminary. Most likely, existing methods are valid in the case of strong connections, whereas subtleties in the analysis come into play in the case of transient, weak, or sparsely-active synapses, and may lead to significant errors as recordings scale up~\citep{Jun2017}. These problems motivate the development of a convincing statistical framework for quantifying the evidence for synaptic interactions beyond the role of background co-modulation.

A fundamental statistical issue, then, is how to detect and quantify the presence of finely-timed spikes in the presence of background fluctuations. The issue becomes delicate when, as has been argued in more general terms~\citep{Brody1999, Marshall2002, Ventura2005, Amarasingham2015, Yu2013}, the background dynamics may be nonstationary and rapidly varying, complicating the preliminary task of accurately estimating their properties. An approach to this problem of estimation has been developed based on a statistical technique called conditional inference~\citep{Amarasingham2012}. The key technical idea is that hypotheses regarding fine temporal structure are formulated in terms of conditional probability distributions on spike times, conditioned on a coarsening of the spike trains. The extent of the coarsening specifies the background timescale. The formulation through conditional distributions allows one to sidestep the estimation problem (for the background dynamics). The statistical technique of {\it jittering} spikes at a certain timescale and then studying the effect of this data-analytic perturbation on the temporal structure of a spike train is an intuitive prototype of this computation. The physiological relevance of this technique was supported by its application in several \invivo findings: most notably, the recovery from \invivo extracellularly recorded spike times of typical \invitro short-term plasticity patterns~\citep{Fujisawa2008}. 

A more precise connection, however, between pairwise spiking data and physiological parameters requires a biophysical interpretation. Several biophysical models of fast monosynaptic dynamics have been proposed, but these largely derive from \invitro observations~\citep{Herrmann2001}. Moreover, none of these models reproduce the extremely fine-temporal spike-spike correlations observed \invivo\citep{Fujisawa2008,English2017} and described above. Here we develop a biophysical model that accounts for these correlations while reconciling millisecond-precise spiking at a synapse and high firing variability. We use the model to demonstrate the use of the conditional approach to inferring fast monosynaptic interactions. Within this framework, we derive procedures for inferring an excess synchrony estimate from spike times, and evaluate its characterization of synaptic properties against the biophysical data.

The work serves to confirm the relevance of the conditional modeling approach, while clarifying its quantitative assumptions, in particular the critical role played by the statistical formulation of a principle of separation of timescale. Challenges associated with biophysical interpretations of the background timescale are highlighted. A related, more general, objective is to contribute to the  important conceptual task of coordinating dynamical and statistical levels of neurophysiological description~\citep{Kass2017}, particularly with respect to the fundamental issue of fast nonstationarities in background dynamics.

\section{Results}

\makeatletter
\setlength{\@fptop}{0pt}
\makeatother
\begin{figure}[!htbp]
\includegraphics[width=\linewidth]{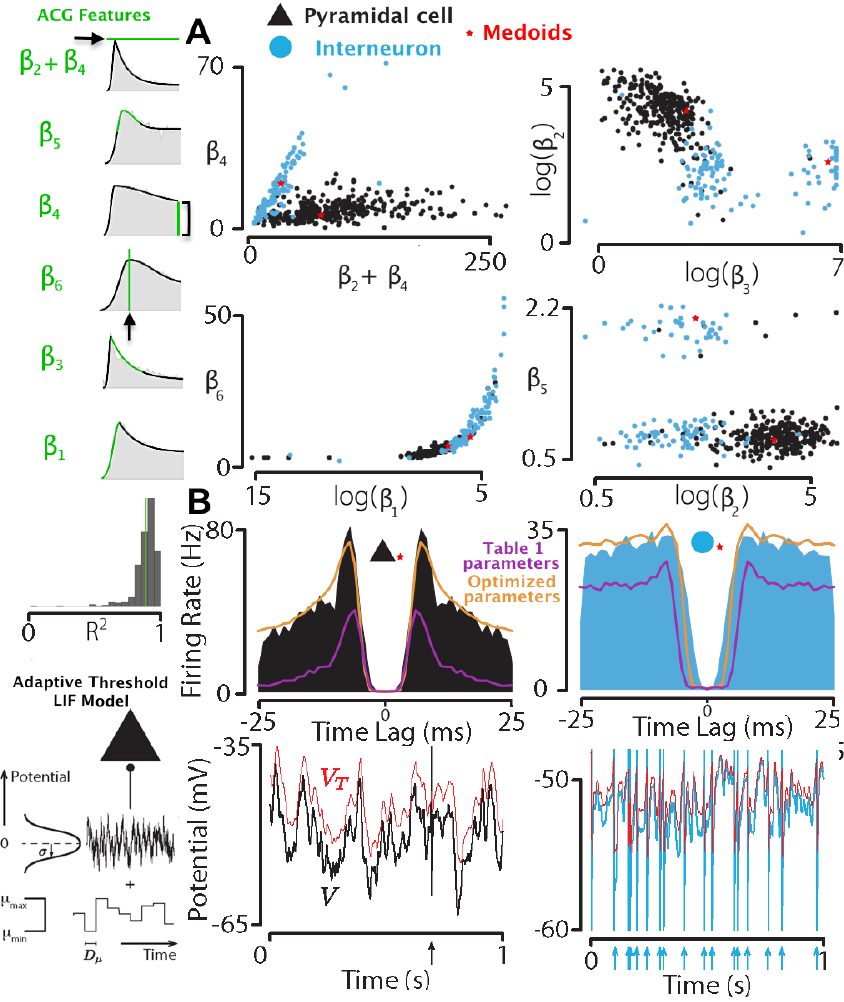}
\caption{{\bf Biophysical models with nonstationary inputs that reproduce \invivo ACG features.} Left insets show a simple sum of exponentials model (see Methods) that was used to parameterize the features of the \invivo auto-correlograms (ACGs). Features determined by the $\beta$ parameters are highlighted in green, example ACGs from the data in gray color, fits in black color. Below these examples, one additional inset shows the distribution of $R^2$ for the model fits (vertical green line represents the average, $R^2$ = 0.89). \textbf{A:} Features of the positive portion of ACGs for CA1 pyramidal and interneurons (data from \citet{English2017}). {\it Top left}: The height of the peak ($\beta_{2}$ + $\beta_{4}$) and baseline firing rate ($\beta_{4}$) (units of Hz). {\it Top right}: The rate of decay away from the peak ($\beta_{3}$) and the difference between the peak and baseline ($\beta_{2}$). {\it Bottom left}: The slope of the rise out of zero lag ($\beta_{1}$) and the lag-coordinate of the peak ($\beta_{6}$). {\it Bottom right}: $\beta_{2}$ and the shape of the peak ($\beta_{5}$). Note that $\beta_{5} \in \{1,2\}$, however the values are wiggled here for visualization. Blue dots and black dots are interneurons and pyramidal cells, respectively, according to the independent labeling method use by English et al., 2017. Red asterisks are medoids of the $\beta$ parameter space determined with the k-medoids (k=2) algorithm, showing correspondence with the cell-type classifications of \citet{English2017}. The medoid is the data point chosen in each cluster to act as its center, and k-medoids seeks to partition the data into k-clusters that minimize the within cluster point-to-medoid distance (see Methods). \textbf{B:} The compatibility of the adaptive threshold LIF model with the \invivo data. The left inset depicts the components of the biophysical model with nonstationary background input $\mu(t)$. {\it Top left}: Black ACG is the pyramidal cell medoid from panels in A (red asterisk in the pyramidal cluster). Orange line is an example ACG simulated from the adaptive threshold LIF neuron with parameters fit to the medoid ACG in the data by global optimization (see Methods). Purple line shows a simulated ACG from the adaptive threshold LIF neuron with parameters from Table 1. {\it Top right}: Blue ACG is the interneuron medoid from panel A (red asterisk in interneuron cluster) with the same fits superimposed. {\it Bottom:} example membrane potential dynamics for each model neuron type with parameters from Table 1 (black and blue lines) as well as the dynamic threshold $V_T$ (red color) for each.}
\label{fig:biophysical-models}
\end{figure}

In the following section, we develop biophysical models that account for the individual and pairwise spiking dynamics \invivoe. First, we describe an integrate-and-fire neuron driven by a stochastic input characterized by fast-timescale nonstationarities. We compare the autocorrelation features of the model with neurons in the \citet{English2017} dataset. Then, a pair of such integrate-and-fire neurons are coupled by common, rapid fluctuations in the statistical parameters describing the stochastic input terms for each neuron. These noisy input processes model the neuron's background input, decoupled from any monosynaptic connections. We then model a monosynaptic connection between the two neurons in two steps. In the first step, we explicitly induce a spike in the postsynaptic neuron a few milliseconds after randomly-chosen spikes in the presynaptic neuron (`injected synchrony'). In the second step, we substitute this injection process with a conductance-based model of the monosynapse. {We compare the CCG features of the conductance-based model neurons with hippocampal neuron pairs recorded \invivoe, which had been labeled as synaptically connected by \citet{English2017}. As already illustrated in Figure 1, the English et al. study partially deconfounds monosynaptic effects from background network activity through juxtacellular perturbation of excitatory neurons in the pyramidal cell (PYR) to interneuron (INT) subcircuit in hippocampal region CA1 of awake mice. The persistence of finely-timed INT spikes following such experimentally-evoked PYR spikes is strongly suggestive of the existence of a monosynaptic connection and, correspondingly, confirms previous `monosynaptic' interpretations of CCG's in spontaneous data~\citep{Csicsvari1998,Fujisawa2008}.

Turning to inference, we formulate the statistical challenge as that of estimating the number of finely-timed spike pairs that can be attributed to a monosynaptic connection, to be used as an indirect proxy characterizing the monosynapse. Effectively, we ask: how much would the peak of the CCG change if the synapse were removed in otherwise identical statistical conditions. This counterfactual is treated as the primary object of estimation~\citep{neyman1923application,rubin1974,lewis1974causation,pearl2009causality,imbens2015}, and its inference can be assessed directly in simulations of the biophysical models. We also similarly address the related issue of synapse detection, and its sensitivity. The inference methods are developed from the injected synchrony point of view, which is of independent neurostatistical interest. The challenge emphasized here is the setting of fast background nonstationarities. To examine this challenging regime, we circumscribe the problem somewhat by making explicit modeling choices. We will defer the discussion of alternative choices, and related issues of interpretation, to the Discussion, where we demonstrate robustness to the main results with relaxations of specifics of the background input assumptions.} 

\subsection{Biophysical models}
Integrate-and-fire neurons provide a minimalist set of mathematically tractable equations that are faithful to cellular physiology~\citep{Gerstner2014},  are commonly used in theoretical neuroscience, and have been repeatedly demonstrated to predict spike times at millisecond precision ~\citep{Gerstner2009,kobayashi2009made,jolivet2008benchmark}. Thus they are natural candidates for  reproducing spike time-based \invivo observations. We constructed these models under the constraint that they quantitatively reproduce the spike auto- and cross-correlograms reported in \invivo studies, and that their parameters are physiologically plausible (see Methods).

The basis of our biophysical modeling is a noise-driven leaky integrate-and-fire neuron with a fast adaptive threshold. Two variables characterize the dynamical state of such a neuron: the membrane potential $V$ and the voltage threshold $V_T$. A spike is emitted whenever $V$ positively crosses $V_T$, which in turn forces $V$ to be reset instantaneously to a fixed value. The dynamics of these two variables are controlled by deterministic equations. While the membrane potential $V(t)$ is constrained by a passive membrane model and the input $I(t)$,
\begin{equation}
\tau_m \frac{dV}{dt} = -V + I,
\label{eq:membrane}
\end{equation}

\noindent the threshold $V_T(t)$ is constrained by a sodium channel model and $V(t)$,
\begin{equation}
\tau_T \frac{dV_T}{dt} = -V_T + f(V),
\end{equation}
$f$ being a linear rectifier function.  The input $I(t)$ is a stochastic process and is intended to mimic the background network activity as seen from the site of spike initiation (see Discussion), 
\begin{equation}
\tau_I \frac{dI(t)}{dt} = -I(t) + \mu(t) + \sigma_I \sqrt{2 \tau_I} \xi(t),
\end{equation}
with $\xi$ a Gaussian white noise process of zero mean and unit variance. As will be apparent later, this background will explicitly exclude the monosynaptic connection between neurons when two integrate-and-fire neurons are modeled as a monosynaptic pair. Details of the biophysical equations can be found in the Methods. 

\subsubsection{Nonstationary firing}
In principle, a neuron's firing rate at time $t$ refers to the instantaneous probability of spiking for specified experimental conditions. This notion is subjective, as it presumably depends on the point of view of an observer with limited information. (For this reason, what we typically refer to as `nonstationary' firing in experimental contexts, is in fact model misspecification, from the point of view of statistics;  \citealp{Harrison2013,Amarasingham2015}.)
 In contrast to the typical experimental situation, here we work with an explicit model for the biophysical dynamics, so we can specify explicitly that the spike train's probabilistic structure derives from lack of knowledge about the signal $I(t)$, which models the background. As a consequence, in the rudimentary setting described above, the model has rich dynamical and statistical structure, but the `firing rate' does not evolve over time (or repeat across `trials'; indeed, there is no trial structure), because the drift term $\mu$ is time-invariant~\citep{Aviel2006, Gerstner2000, Ostojic2011, Gerstner2014}. Our theoretical perspective in this study borrows from that of nonparametric statistics in that $I(t)$ will be intuitively chosen to maximally promote finely-timed (`apparently monosynaptic') spike pairs among a broad class of such background models, in the absence of a monosynaptic connection.

To give `nonstationarity' an explicit meaning, let us consider the setting where $\mu(t)$ is itself a rapidly-varying stochastic process, as a model of {\it in vivo} trial-to-trial variability. That is, in the context of this particular model, we are interested in studying the regime where $\mu(t),$ which is treated as unknown in the inference problem, varies quickly enough to challenge standard approaches (such as derived from maximum likelihood) to neurostatistical inference. Here the inference of interest is of monosynaptic properties. Such a model is in the spirit of `doubly-stochastic' spiking models \citep{Churchland2012} (but, again, see \citealp{Amarasingham2015} for warnings about this language.)\footnote{To be explicit about the subtlety in the setting here: in fact, we will model $E[\mu(t)]$ as time-invariant (Eq. \ref{piecewise-constant}). Thus, for an observer that does not know $\mu(t)$, the spike probabilities are time-invariant. On the other hand, for an observer that knows $\mu(t)$, the spike probabilities are not time-invariant.} {A natural choice in this direction is to introduce a stochastic $\mu(t)$ that fluctuates more slowly than the Gaussian colored noise.} Motivated by simplicity, we model $\mu(t)$ as piecewise constant on equal-length intervals with random independently- and uniformly-distributed amplitudes,
 \begin{equation} \label{piecewise-constant}
 \mu(t) = \displaystyle \sum_{i=1}^{m} \mu_i \mathbbm{1} \big\{ (i-1) D_{\mu} \leq t < i D_{\mu} \big\},
 \end{equation}
 where $\mathbbm{1} \big\{A\}$ represents  the indicator function of $A$ (i.e., taking the value 1 when the expression $A$  is true, and 0 otherwise).
 The length of the intervals, $D_\mu$, models the timescale of this source of spiking variability. $\mu_1,\mu_2,...,\mu_m$ are independent and uniformly-distributed (see Methods for details). For now, we view a realization of $\mu(t)$ as an approximation to a smooth function. The motivation for this form of approximation is related to details of the statistical tools we will use (e.g., `interval jitter' and related tools, to be described below) for inference, and will simplify the biophysical-statistical correspondence we seek to examine. Later, we will relax this assumption and find that our principal conclusions are robust to the piecewise constant form, using noise processes $\mu(t)$ with continuous realizations, or with variable $D_{\mu}$.
 
 \subsubsection{Fitting the biophysical model to data}

 {To demonstrate the correspondence of this integrate-and-fire model with the \invivo  data of \citet{English2017},
 we first developed a simple sum of exponentials model (Eq.~\ref{eq:acgshapemodel}, see Methods) to fit the shape of the positive portion of auto-correlograms (ACGs) for pyramidal neurons and interneurons (see left insets in Figure~{\ref{fig:biophysical-models}A} for sample fits). The purpose of using this model was to reduce dimensionality in comparing the \invivo ACGs with the biophysical ACGs by way of features that have intuitive meaning. (The utility of simple methods such as averaging different ACGs per bin are limited because specific features, such as the ACG peak, occur at different lags in different ACGs.) We found the shape of the ACGs can be well-approximated by two exponential functions tied together at the ACG peak (average $R^2 = 0.89$, see Figure~{\ref{fig:biophysical-models}}).  In addition to quantifying the features of the ACGs, the procedure classifies the neuron types (the colors in Figure~{\ref{fig:biophysical-models}A} are based on the independent labeling method of  \cite{Sirota2008}), further corroborating the model's validity. Figure~{\ref{fig:biophysical-models}A} shows the features of pyramidal cells and interneurons for the whole dataset, and the medoids of the pyramidal and interneuron clusters (see Methods).
 
Using global optimization methods, we optimized the parameters of the integrate-and-fire model to fit the medoid ACGs in the data (Figure~{\ref{fig:biophysical-models}B}). Unsurprisingly, there were various parametrizations discovered across independent runs that fit the medoid ACGs well. The statistics ($mean \pm std$) of the optimization solutions across ten independent runs were: for the pyramidal neuron, $\tau_T = 13.54 \pm 1.33$ ms, $V_i = -52.80 \pm 4.81$ mV, $\tau_I = 11.76 \pm 5.03$ ms, $\mu_I = -47.39 \pm 6.55$ mV, $\sigma_I = 14.72 \pm 5.71$ mV, $D_{\mu} = 18.97 \pm 1.01$ ms, refractory $= 2.45 \pm 1.2$ ms, $(\mu_{max}-\mu_{min})/2 = 16.44 \pm 6.32$ mV, $\alpha = 0.45 \pm 0.32$, $\tau_m = 20.06 \pm 10.14$ ms, and $V_t - V_i = 7.84 \pm  5.21$ mV and, for the interneuron, $\tau_T = 0.88 \pm 0.09$ ms, $V_i = -59.27 \pm 3.14$ mV, $\tau_I = 4.97 \pm 3.26$ ms, $\mu_I = -55.04 \pm 1.06$ mV, $\sigma_I = 2.93 \pm 0.96$ mV, $D_{\mu} = 4.94 \pm 2.16$ ms, refractory $= 2.26 \pm 0.68$ ms, $(\mu_{max}-\mu_{min})/2 = 11.09 \pm 1.89$ mV, $\alpha = 0.74 \pm 0.04$, $\tau_m = 19.59 \pm 8.95$ ms, and $V_t - V_i = 1.57 \pm  1.06$ mV. These averages were used for the ``optimized'' parameters in Figure~{\ref{fig:biophysical-models}}B and also for the bottom panels for Figure~{\ref{fig:monosynapse-estimation}}A-C. Except, for the latter, the following parameters were further qualitatively tuned so that the CCG Sharpness approximately matched the average \invivo CCG Sharpness curve for monosynaptically connected pairs seen in Figure~{\ref{fig:monosynapse}}: for the interneuron, $\tau_T = 0.22$ ms, $\alpha = 0.77$, and for both neurons, $D_{\mu} = 10$ ms, $\mu_I = -51.21$ mV, and $(\mu_{max}-\mu_{min})/2 = 10.7$ mV.

These optimizations serve two purposes. Firstly, we regard them merely as an existence proof that the integrate-and-fire model with adaptive threshold can be tuned to sharply fit the ACG features of the \invivo data on timescales on the order of 50 ms. Second, in Figure~{\ref{fig:monosynapse-estimation}}, we use them with a another set of parameters comparatively to explore the relationship between the biophysical parameters and assumptions of the statistical model. 

This other set of parameters (Table 1) are constrained more by empirical and theoretical considerations, while also taking into consideration the optimizations. In this set, the passive membrane parameter values were chosen following~\citet{Ostojic2009}, along with the monosynapse parameter values (the latter for Table 1 and Table 2).  The input's time constant, $\tau_I$, was chosen based on theoretical and empirical considerations~\citep{Brunel2001, Fourcaud2002, Gerstner2014}. We tuned it large enough compared to the membrane time constant so that the input $\mu(t)$ had a strong impact on the instantaneous spike rate; but small enough so that it affected as little as possible the input variance on the membrane timescale (for comparison, $\tau_I \sim 0$ in~\citealp{Ostojic2009}, as the authors studied a white noise input). The active threshold parameter values were chosen according to the experimental values~\citep{Fontaine2014, Mensi2016}, taking into account the underlying biophysical constraints~\citep{Platkiewicz2011}, which agreed quite well with the optimization results. The refractory period was not taken into account for Table 1.

\begin{figure*}[!htbp]
\includegraphics[width=\linewidth]{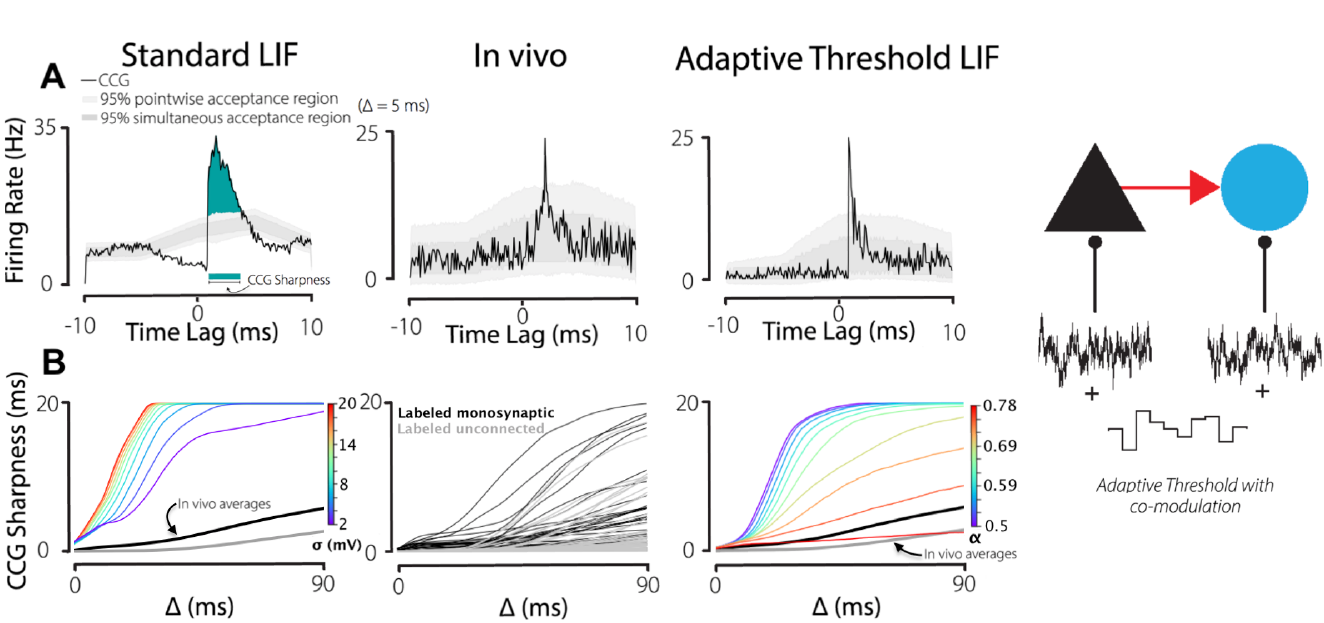}
\caption{{{\bf Modeling the in vivo monosynaptic interaction.} \textbf{A:} Example cross-correlograms (CCGs), with acceptance bands specified by 5 ms interval jitter tests. {\it Left:} The CCG from a standard (constant threshold) LIF neuron with colored noise. Falling outside the acceptance bands indicates spike timing on a finer than 5 ms timescale. Light gray bands accommodate multiple hypothesis testing. See Methods for a description of the technical meaning of these statements.  {\it Center:} the CCG between a CA1 pyramidal cell and interneuron in vivo \citep{English2017}. {\it Right:} the LIF model with adaptive threshold dynamics. In the standard LIF model, parameters were hand-tuned to make the CCG peak nearly as sharp as possible ($\sigma_{I} = 2$ for the postsynaptic neuron and $\mu_{I} = -51.5$ mV to approximately equalize the firing rates across the examples). Otherwise, the parameters are the same as those in Table 1. The adaptive threshold parameters in this example were also chosen to push the model towards limiting behavior for CCG sharpness ($\tau_{T} = 0.22$ and $\alpha=0.78$ for the postsynaptic neuron). Remaining parameters are the same as for the constant threshold LIF example CCG (specified in Table 1). \textbf{B:} Comparing CCG sharpness systematically. We define CCG sharpness as the maximum width of the peak exceeding the simultaneous acceptance  band at a $\Delta$-level interval jitter test (see example in leftmost panel A for $\Delta = 5$ ms). {\it Left}: The sharpness of the CCG peak in the standard LIF model is notably parameterized by the postsynaptic Gaussian noise ($\sigma_{I}$). The sharpness of the peak relative to many thresholds with $\Delta \in \{0.1,0.2,...,0.89,90\}$ ms  and $\sigma_{I} \in \{2,4,...,20\}$ mV is shown. {\it Center}: Analogous curves for all pyramidal-to-interneuron pairs in the English et al. data set. Note that with the standard LIF model we could not reproduce the \invivo CCG peak. Right panel: In the adaptive threshold model, the CCG becomes sharper as the postsynaptic $\alpha$ increases, here plotted is $\alpha \in \{0.5,0.53,...,0.78\}$. Note that the in vivo curves (from center panel) are averaged pointwise and plotted in the left and right panel for comparison.}}
\label{fig:monosynapse}
\end{figure*}

\subsubsection{Co-modulated firing}
Thus far we have discussed single neuron dynamics, but microcircuit connectivity inference involves the analysis of at least a pair of spike trains. Two neurons that are not monosynaptically connected can emit spikes that appear correlated on a millisecond-timescale even if the underlying coupling mechanism operates on a slower timescale~\citep{Brody1999, Ventura2005, Yu2013}. We propose a simple mechanism of firing co-modulation. The circuit is composed of two integrate-and-fire neurons, as just described, but the piecewise constant input is now common to (i.e., exactly the same in) both neurons. The two Gaussian colored noise processes remain independent. We use the same parameters we used previously for reproducing pyramidal and interneuron ACG's. On broad timescales, the CCG exhibits a hump-shaped envelope typical of slow co-modulation.

\begin{figure}[!htbp]
\includegraphics[width=\linewidth]{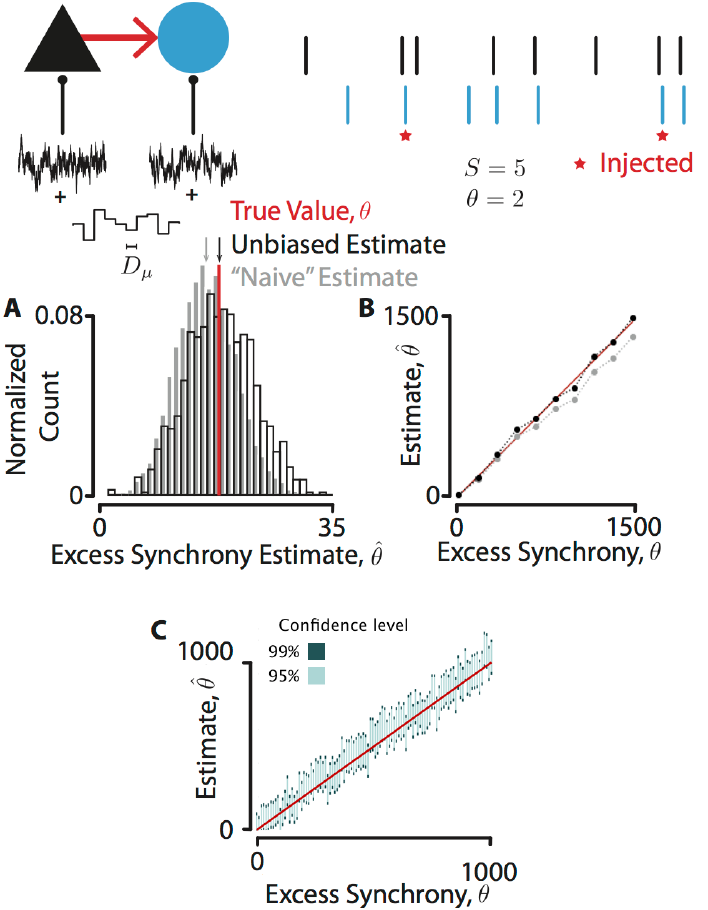}
\caption{{\bf Estimation of the excess synchrony (artificial spike injection).} In these numerical experiments, we implemented only the synapse model with artificial spike injection. \textbf{A:} First, we generated multiple spike train pairs using the biophysical model ($1,000$ trials of $1,000$ s duration), but kept the number of injected spikes the same across trials (solid red line; $\approx 18$ spikes). We estimated the amount of injected spikes for each trial using both our closed-form formula (black), and a subtraction of the jitter-corrected synchrony count from the total synchrony count (gray), taking the jitter interval length equal to the piecewise constant input's interval length. The arrows in the corresponding colors indicate the average of each distribution. Note that the closed-form estimate is closer to the true value compared to the standard jitter-corrected estimate. \textbf{B:} Similar numerical experiment ($10$ trials of $100,000$ ms), analysis, and color code, but the amount of injected spikes was varied across trials. The solid red line indicates the first diagonal. Note that the discrepancy between the naive estimate and the true count increases with the true count, while it remains the same for our proposed unbiased estimate. \textbf{C:} Similar numerical experiment to the one performed in B, but with 1,000 trials for $\theta \in \{1,2,...,1000\}$. For each $\theta$, $95\%$ and $99\%$ general confidence intervals were computed (only 100 equally-spaced confidence intervals are plotted as vertical lines over the true parameters). Across all 1,000 trials, the measured coverage percentages were 93.3\% coverage for $95\%$\;CI's, and 98.3\% coverage for 99\%\;CI's.}
\label{fig:injected-synchrony}
\end{figure}

\subsubsection{Ultra-precise monosynaptic spike transmission from input conductance dynamics}
Our most complete biophysical model of a monosynaptically connected cell pair is based on the co-modulated firing mechanism, but the target neuron receives an additional input. This input is classically described as  an input current $g_{s} (E_{s}-V)$ to the membrane equation (Eq.~\ref{eq:membrane}) , where $g_{s}$ is the synaptic conductance, and $E_s$ is the synaptic reversal potential. More precisely, each presynaptic spike increments $g_s$ by a fixed amount, following a fixed latency delay, which then decays exponentially to its baseline level,

\begin{equation}
\tau_s \frac{dg_s}{dt} = -g_s.
\end{equation}

As a result, the dynamics of $g_s (E_{s}-V)$ following a single presynaptic spike mimics the typical trajectory of a postsynaptic current (PSC).  See Methods for more details. 

{The main motivation for using the adaptive threshold model is to reproduce the extremely sharp CCG found \invivo that \citet{English2017} validate as a monosynaptic signature. To motivate this modeling decision, we compare the CCG sharpness of monosynaptically labeled pairs in their data, with both the standard leaky integrate-and-fire (LIF) model with colored noise (constant threshold) and with the adaptive threshold model (Figure~{\ref{fig:monosynapse}}).

We used the upper simultaneous acceptance band derived from a $\Delta$-jitter hypothesis test (see Methods) as a threshold to distinguish significant peaks from slow timescale statistical fluctuations. These tests are designed to separate fine-timescale 
from coarse-timescale spike interactions \citep{Amarasingham2012}; the separation of timescale is parameterized by $\Delta$. We make comparisons over a range of $\Delta$. For a fair comparison, we set the parameters of the constant threshold LIF model so that the CCG was nearly as sharp as possible while keeping other parameters the same as in Table 1. In the constant threhold LIF, three ways the CCG peak can be made more sharp are i) by increasing the peak synaptic conductance ($g_0$) or ii) decreasing the synaptic decay time ($\tau_s$), or iii) decreasing the amplitude of the postsynaptic gaussian noise $\sigma_I$ \citep{Ostojic2009,Herrmann2001}. We show the CCG sharpness for all  $\Delta<100$ ms for a range of $\sigma_I$ (Figure~{\ref{fig:monosynapse}}A) with fixed $g_0$ and $\tau_s$. Figure~{\ref{fig:monosynapse}}B shows the same CCG sharpness curves for every neuron pair in the \citet{English2017} data set. Note that the sharpness of neuron pairs labeled monosynaptic tends to increase more gradually as $\Delta$ increases than those not labelled monosynaptic, indicating that, at least with respect to the labeling methods of English et al., monosynaptic interactions are marked by very fine timescale spike relationships. In these terms outlined above, we could not reproduce the extreme temporal precision of the \invivo monosynaptic interaction using the standard constant threshold LIF model. 

Using the same values for $g_0$ and $\tau_s$ and with $\sigma_I = 2$ mV (which yielded the sharpest CCG peak in Figure~{\ref{fig:monosynapse}}A), the same curves are plotted for various parameters of the postsynaptic adaptive threshold model (Figure~{\ref{fig:monosynapse}}C). The coincidence detection property of the interneuron increases as $\alpha$ increases and $\tau_T$ decreases. The curves are remarkably similar to those generated by the monosynaptically labeled pairs \invivoe. In this sense, the model appears to capture the basic temporal features of the monosynaptic interactions well (refer to the Discussion for commentary on alternative mechanisms).

\subsection{Estimating excess synchrony: separation of timescale and injected synchrony}

{In what follows, the models developed above will serve as ground truth for studying the statistical problem of monosynaptic inference in nonstationary conditions.} For the purpose of inference, we develop a statistical model for a monosynaptically connected pair of spike trains that dissociates postsynaptic spike train structure into coarse and fine timescale components. The fine timescale component induces fine-timescale correlations between the pre- and postsynaptic spike trains, and will be associated with monosynaptic phenomena. The coarse timescale component effectively models the influence of unobserved synaptic inputs (the `background'), and is designed in mind of the the concern that the background process is confounded by nonstationary effects. We refer to this distinction between fine and coarse timescales as the separation of timescale principle.

\subsubsection{Formulation}
\label{injectedformulation}
The postsynaptic train is modeled as a superposition of two trains: a {\it background} spike train ($\vect B$) and an {\it injected} spike train ($\vect I$).  The statistical properties of the background train are modeled nonparametrically to accommodate nonstationarity. Technically, the nonparametric model is that the conditional probability distribution on the background spike train is uniform, conditioned on knowledge of the reference train and on knowledge of the spike counts of the target train in intervals of length $\Delta$. $\Delta$ parameterizes the characteristic timescale of the background spike trains. We make this precise as follows.

A spike train, $\vect T$ for example, is modeled in discrete time, and we write $\vect  T = (T_1, \ldots, T_{|\vect T|} )$, where $T_i$ is the time of the $i$\textsuperscript{th} spike in $\vect T$, and $|\vect T|$ is the number of elements in $\vect T$ (in places where $A$ is not defined to be a set, $|A|$ retains the conventional meaning of absolute value). Time is partitioned into equal-length intervals of size $\Delta$, and the spike counts in $\vect T$ in the intervals are denoted $\vect  N(\vect T) = (N_1(\vect T), \ldots, N_m(\vect T))$ where $m$ is the number of intervals~\citep{Amarasingham2012}. 

Given two spike trains $\vect R$ and $\vect T$, their synchrony count is defined as, \begin{equation}
S = \displaystyle \sum_{i=1}^{|\vect R|} \sum_{j=1}^{|\vect T|} \ind \{T_j-R_i= \textrm{lag}\}
\end{equation}
where $\textrm{lag} > 0$ represents the synaptic delay in units of time bins. This counts the number of (lagged) synchronies between $\vect R$ and $\vect T$. In practice, one of the pair of spike trains can be shifted in time so that the lag can effectively be treated as 0, which is the practice we adopt here.

Thus the time discretization specifies the resolution of synchrony. Here the trains $\vect R$ and $\vect T$ will be referred to as the reference and target, respectively. In addition, we impose the condition that all intervals that contain a spike in $\vect T$ also contain a spike in $\vect R.$ (Spikes in the target train that do not satisfy this condition are not in $\vect T$; they do not contribute to the synchrony count by definition so they can be ignored without loss of generality.) 



\subsubsection{Injected synchrony model}
When evaluating a cell pair that is putatively monosynaptic, the candidate presynaptic train is chosen as reference and the candidate postsynaptic train as target. In the following, we will assume that the reference train is fixed and the target train is random. (Technically, the probability distribution on $\vect T$ is in fact the conditional probability distribution on $\vect T$, given $\vect R$.). The model (``injected synchrony'') is that (conditioned on $\vect R$), the target train $\vect T$ is a superposition of two point processes (spike trains): an ``injected'' train, $\vect I$, and a ``background'' train, $\vect B$. $\vect B$ is assumed to be conditionally uniform, conditioned on $\vect N(\vect B)$ and $\vect R$. In other words, conditioned on $\vect N(\vect B)$ and $\vect R$, all spike trains (that are consistent with $\vect N(\vect B)$ and $\vect R$) are equally likely outcomes for $\vect B$:
\begin{equation} \label{eq:cond-unif}
    P( \vect B=\vect b | \vect N(\vect B)=\vect N(\vect b),\vect R=\vect r ) = c( \vect N(\vect b), \vect r),
\end{equation}
for some function $c(\cdot,\cdot)$. (In yet other words, the conditional likelihood that $\vect B$ takes the value $\vect b$, given knowledge of $\vect N(\vect B)$ and $\vect R$, depends on $\vect b$ only through $\vect N( \vect b)$.)\footnote{No constraints are placed on the probability distribution of $\vect N(\vect B)$ itself, and this is the source of robustness to nonstationarity, broadly defined. This conditional modeling framework for modeling temporal structure in spike trains is motivated and developed from several points of view in prior work  \citep{Amarasingham2012,Amarasingham2015,Harrison2014,Harrison2013}. See \cite{Amarasingham2012} for a thorough introductory exposition. As an example, if $\vect B$ is conditionally a homogeneous Bernoulli process, conditioned on $\vect R$, then Eq. (\ref{eq:cond-unif}) is satisfied. (The homogeneous Benoulli process approximates the homogeneous Poisson process in discrete time.) But the model is far broader than this. For example, $\vect N(\vect B)$ can be deterministic and Eq. (\ref{eq:cond-unif}) can still be valid (cf., Section A.2 of \cite{amarasingham2011conditional}). } Note that two spikes can be in the same bin in the superposition, which is an approximation. All of the spikes in $\vect I$, on the other hand, are assumed to be synchronous with $\vect R$. The synchrony is delayed by the parameter $\textrm{lag}$, meaning that all spikes in $\vect I$ occur $\textrm{lag}$ time bins after a spike in $\vect R.$\footnote{For intuition, a canonical example can be constructed by generating two distinct independent, background spike trains $\vect B_1$ and $\vect B_2$, which are conditionally uniform, conditioned on $\vect N(\vect B_1)$ and $\vect N(\vect B_2)$, and superposing a homogeneous Bernoulli process synchronously (with appropriate lag) onto both trains (see \citealp{Amarasingham2012}, for example, for examples based on Cox processes for the background.)}       
Given the model, the inference problem is to estimate the number of `injected' synchronies, $\theta=|\vect I|$. $\theta$ and ${\vect R}$ are fixed, and, for now, we assume that the background timescale $\Delta$ is known (cf. Figure~{\ref{fig:relaxbackground}}). In effect, the inference problem asks: what is the excess synchrony, assuming there is a background process that has timescale $\Delta$, and the excess synchronies are exactly timed (measured at the resolution of discretization)?

The model induces a decomposition of the total synchrony, $S$ as,

\begin{equation}
S = S_{\Delta} + \theta,
\label{eq:synchrony-balance-results}
\end{equation}
where the synchrony count $S_{\Delta}$ are synchronous pairs composed of background spikes (sometimes called chance or accidental synchrony), and the synchrony count $\theta$ is the number of injected spikes (sometimes called excess or non-accidental synchrony) [see Methods]. $S_\Delta$ and $\theta$ are latent (unobserved) random variables.

\subsubsection{Interpretation} \label{sec:IS-Interpretation}
Thinking in terms of neurophysiology, the essence of this model is the assumption that a spike (in the target train) can be unambiguously labeled as caused by {\it either} a slow (background) process, {\it or} a fast synchrony-inducing process. The statistical problem is to separate the fast synchrony ($\theta$) from the slow synchrony ($S_\Delta$) without access to the labels. The timescales of fast and slow are quantitatively modeled through the characteristic timescale $\Delta$. Correspondingly, the model breaks down when such a causal attribution is ambiguous (and would likely require, among other components, a multiscale framework), {as for example, in the case of weak synapses} (see Discussion). This is the sense in which the model expresses a `separation of timescale' hypothesis for monosynaptic characterization. 

The injection model in the case $\theta=0$ is in fact the null hypothesis for the {\it interval jitter} hypothesis testing procedures successively elaborated in precursor work \citep{Amarasingham2012,Date1998}. Statisticians commonly advise that tests of hypotheses be accompanied by an assessment of magnitude of effect, to distinguish `scientific' significance from `practical' significance, among other reasons \citep{wasserman2013all}. Injected synchrony can be motivated as one such measure for interval jitter hypothesis tests. This motivation has other applications in neurophysiology (see \citealp{martin2015spike,shahidi2019high,barbosa2020interplay} for precise examples).

\subsubsection{Point estimation of $\theta$} 

Consider a spike train pair generated by the injected synchrony model. An implication of the conditional uniformity assumption for the spike train $\vect B$ is that one can randomly translate the spikes in $\vect B$ without changing its statistical properties, so long as one {preserves $\vect N(\vect B)$,} the spike counts in the intervals. This is one basis for the interval jitter hypothesis tests referred to above. In practice, this can be done by sampling from the uniform distribution on the set of spike trains consistent with $N(\vect B)$. When used to generate surrogate data for hypothesis tests, this procedure is called {\it interval jitter}, because the original spikes in $\vect B$ are, effectively, {\it jittered}.\footnote{Note that i) these jitter perturbations are distorted slightly via the role of the interval positions, ii) those distortions are necessary \citep{Platkiewicz2017}, and also iii) such tests can in principle be performed analytically, without surrogate-generation.} In correspondence to this computation, a natural, naive estimate of the amount of injected synchrony $\theta$ could consist in jittering all the spikes, computing the resulting synchrony, and subtracting it from the original (observed) synchrony. Variations on such a ``jitter-corrected'' synchrony estimate have been implemented in the literature, although described in less formal terms \citep{Amarasingham2012,martin2015spike,smith2008spatial,Fujisawa2008,barbosa2020interplay}. Intuitively, there is some sense to this procedure. Informally, the mean of the resulting jitter-derived distribution furnishes an estimate of $\theta$, and, speaking loosely again, the spread of that distribution would provide some measure of variability. However, it is not hard to see, as clarified by the separation of timescale model here, that this would introduce bias into the estimate, since the logic of jitter resampling only applies to the background spikes. Presumably, a more appropriate procedure would jitter only the spikes in $\vect B$. But, without access to the {\it labels} which identify which target spikes derive from $\vect B,$ and which derive from $\vect I$, it is not possible to distinguish an injected spike from a background spike. The naively-resampled injected spikes would contribute to the synchrony counts in the jitter-derived distribution only, thus corrupting the estimate of $\theta$.

To make more precise sense of the standard approach, put a probability distribution on the possible labelings identifying the source of the target spikes. Jitter-correction then arises from a specific assumption about the distribution on possible labelings, but without bias correction. Make the assumption that, given knowledge of the reference train and treating $\theta$ as an unknown but fixed parameter, all consistent labelings of the target train are equally likely (all labels are equally likely, `LEL')

Then define
\begin{equation}
\bar{r} := \frac{1}{|\bm T|} \sum_{i=1}^{|\bm T|} N_{r,i},
\end{equation}
where $N_{r,i}$ specifies the number of spikes in the reference train in the interval containing the $i$'th target spike, and ${|\bm T|}$ is the number of spikes in the target train. In the Methods section (Section \ref{sec:injected-synchrony-model}), , we show that it then follows that, under the LEL assumption,
\begin{equation}
\hat{\theta} = \frac{ S - \frac{\bar{r}}{\Delta}|\bm T| }{1-\frac{\bar{r}}{\Delta}}
\label{eq:injection-estimate}
\end{equation}
is an unbiased estimator of $\theta$, meaning that:
\begin{equation}
E[\hat{\theta}]=\theta.
\end{equation}
$\hat{\theta}$ can thus be seen as a bias-corrected analogue of `` jitter correction'' as described above. (Jitter correction is a Monte Carlo approximation of $S - \frac{\bar{r}}{\Delta}|\bm T|$.)

Even more relevant to the problem at hand, in the specific case that the number of spikes in the reference train is constant is non-empty intervals, unbiasedness holds independently of the LEL assumption (cf., the main theorem in \ref{sec:injected-synchrony-model}); the latter situation is particularly relevant in the case of sparse firing, where sparseness is defined relative to $\Delta$), {and which is the dominating regime in the experiments reported here. With the biophysical parameters of Table 1 where $\Delta = 10$ ms, the $N_{r,i}$ occur with the following frequencies: $97.4\% = 0$ spikes,
$2.38\% = 1$ spikes,
$0.18\% = 2$ spikes, and
$0.002\% = 3$ spikes.} The formula emphasizes the role of $\Delta$ in the estimation of $\theta$.

\begin{figure*}[!htbp]
\includegraphics[width=\linewidth]{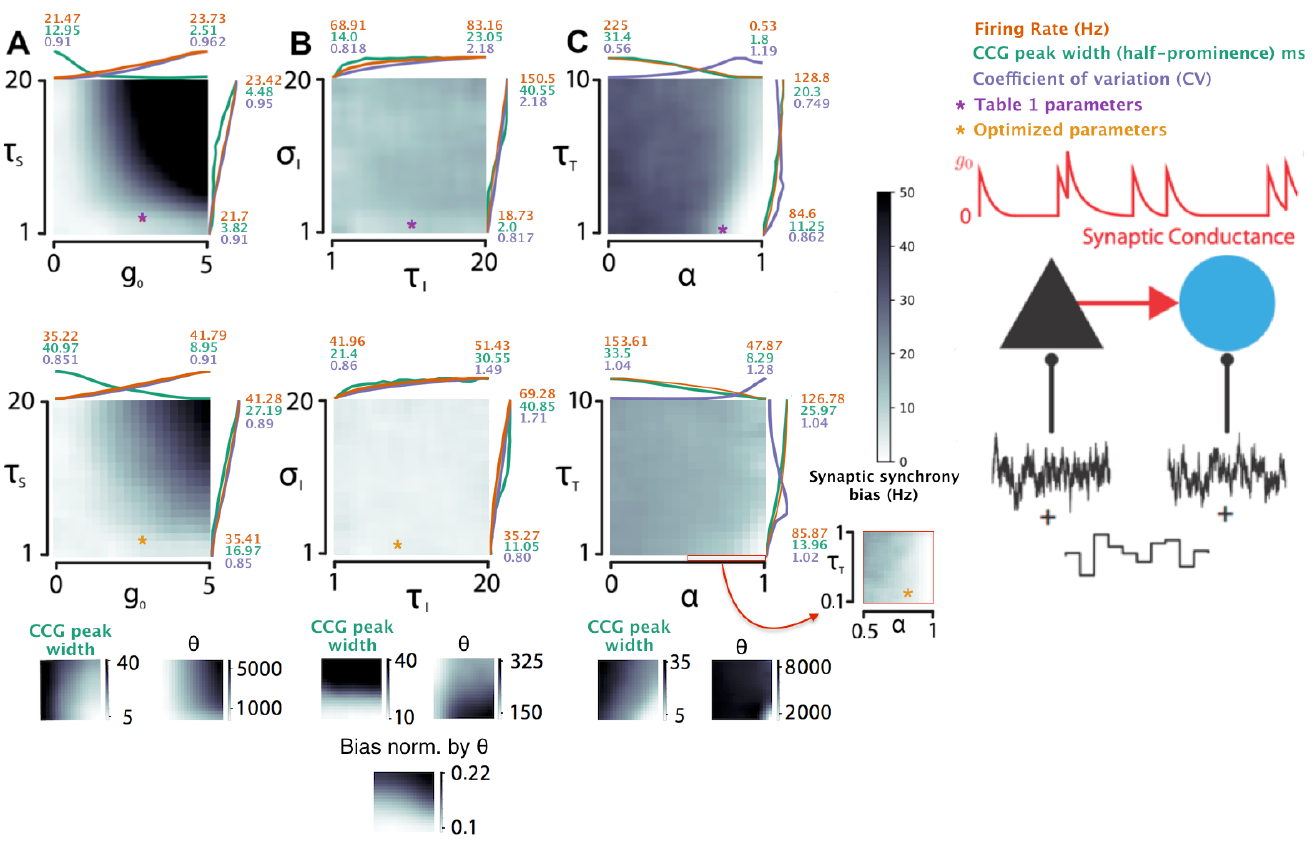}
\caption{{\bf Identification of synaptic synchrony rate.} The empirical bias of the (conditional) synaptic synchrony rate, $|\theta - \hat{\theta}|/(|\textbf{R}|\cdot 10^{-3})$ (Hz), studied as a function of various postsynaptic model parameters in the synaptic conductance model. Parameters other than those being varied, are either fixed at the values in Table 1 (top row) or fixed at the average values obtained in the global optimizations of Figure~{\ref{fig:biophysical-models}} (with a few caveats described in the main text). Each parameter took on twenty equally-spaced values in the intervals described in the following sub-descriptions.  \textbf{A:} Peak synaptic conductance ($g_0 \in [0,5]$ $nS$) and synaptic decay ($\tau_s \in [1,20]$ ms). \textbf{B:} Timescale ($\tau_I \in [1,20]$ ms) and standard deviation ($\sigma_I \in [1,20]$ mV) of the postsynaptic independent noise. \textbf{C:} The adaptive threshold parameters, ($\tau_T \in [1,20]$ ms and $\alpha \in [0,1]$). Additionally, for the optimized parameters, a finer resolution grid is shown with $\tau_T \in [0.1,1]$ ms and $\alpha \in [0.5,1]$). The firing rate, coefficient of variation, and CCG half-width are averaged for the corresponding rows and/or columns of the grids (colored plots adjacent to grids). For ease of visualization, each of the three individual curves is normalized to a common range; the unnormalized values at the endpoints are provided for reference (color-coded). The colorbar is consistent for all plots. Purple and orange asterisks show the biophysical parameters at the values they are fixed at in simulations where they are not being varied. For the optimized parameter set in the bottom panels, insets show the corresponding grids of $\theta$ and CCG half-width. For the middle panel, the synaptic synchrony bias is shown, but normalized by $\theta$ instead of $|\textbf{R}|$ in the inset (i.e., $|\theta - \hat{\theta}|/\theta)$}.
\label{fig:monosynapse-estimation}
\end{figure*}

\subsubsection{Validating the theoretical predictions}
\label{sec:validation-estimation}


To validate our proposed estimation approach (Eq.~\ref{eq:injection-estimate}), we first used the biophysical model with monosynaptically injected spikes. The pyramidal model is taken as presyanptic, and the interneuron as postsynaptic (see Figure~{\ref{fig:biophysical-models}} and Table 1). For each randomly chosen presynaptic spike, we create a postsynaptic spike delayed by a given latency. We call this spike pair an ``injected'' synchrony. Although this is quite a strong simplification of biophysics, it provides a precise demonstration of unambiguous, causal monosynaptic transmission that is independent of the background input. 

First, we simulated numerous trials with this model (1,000 trials of $1,000$ s duration), injecting the same number of synaptic spikes in each trial ($\approx 18$ spikes). For each trial, we estimated the amount of excess synchrony using both our formula and the standard jitter-corrected subtraction. For the jitter interval length ($\Delta$), we used the interval duration of the piecewise constant input ($D_{\mu}$). We represented the empirical distribution for both estimates in Figure~{\ref{fig:injected-synchrony}A}. The distribution and mean of the unbiased estimate is centered around the true injected count, unlike the standard jitter-corrected estimate, providing some confirmation that the statistical injected synchrony model can plausibly be mapped onto an integrate-and-fire neural circuit -- the background firing produced by the LIF-based model is well-captured by the jitter null here. 

In the next experiment, we replicated the previous one (with $10$ trials of $100,000$ s duration), but injected a different number of synaptic spikes in each trial (ten equally spaced values ranging from $0$ to $1,500$ spikes). For each trial, we computed the unbiased and naive estimates (Figure {\ref{fig:injected-synchrony}B}). The results confirm the observations made in the previous experiment. It also shows that the unbiased estimate is close to the true value for sufficiently long trials. The bias of the naive estimate increases as the true value increases, which can be understood as a consequence of jittering all spikes in naive estimation.

\subsubsection{Interval estimation of $\theta$}

$\hat{\theta}$ can vary substantially around $\theta$, especially for highly nonstationary firing and short time windows. This motivates the construction of more general, exact confidence intervals to calibrate this variability. As outlined in Methods, the basic idea is to compute lower and upper bounds for the distribution of synchrony under the appropriate conditioning, and then to use these bounds to construct confidence intervals for $\theta.$ The bounds can be computed over all possible labelings, so that the confidence intervals make no assumptions about the distribution over labellings, and are exact. Using this technique, we conducted a similar experiment to Figure~{\ref{fig:injected-synchrony}}B, but for each $\theta$ we computed $95\%$ and $99\%$ general confidence intervals (Figure~{\ref{fig:injected-synchrony}}C). In $1,000$ trials each of duration $100,000$ ms  we injected spikes in each trial $\theta \in \{1,2,...,1000\}$. Then, for each $\theta$, $95\%$ and $99\%$ confidence intervals were computed. Across all 1,000 trials, the 95\% confidence intervals had 93.3\% empirical coverage, and the 99\% confidence intervals had 98.3\% empirical coverage.

{
\subsubsection{Monosynaptic identifiability in a conductance model}
\label{sec:biophysics-monosynapse}
Thus far, we have validated the statistical predictions of the injected synchrony model on the data generated by our biophysical model with monosynaptically injected spikes.  The integrate-and-fire models were parameterized to give rise to ACG's and CCG's that quantitatively fit the central tendency of those observed \invivoe. This suggests that the background processes generated using integrate-and-fire models are modeled by the corresponding interval jitter null hypothesis with some fidelity. To examine this correspondence with more realistic synaptic mechanisms, we implemented the synaptic conductance model. In this section, we study the bias of $\hat{\theta}$ numerically, as a function of various biophysical parameters to explore the correspondence between the injected synchrony model and neuronal dynamics.

Moving to a conductance-based synapse model, defining the value for $\theta$, the number of spikes attributable to the monosynapse in a biophysical simulation, is necessarily more subtle. While a synapse may influence its postsynaptic target in a variety of ways, the injected synchrony model is inspired by an implicit neurophysiological interpretation of the very thin CCG peak found \invivo (Figure~{\ref{fig:monosynapse}}). Accordingly, we isolate the lag with the most total observed synchronous pairs (i.e., where  the CCG peaks), and we define our problem as that of inferring the number of synchronous spike pairs, at that lag, which are {\it caused} by the monosynapse. This number is defined operationally as the difference between the peak-lag synchrony observed in simulation, and the `counterfactual' peak-lag synchrony observed in simulation with the monosynapse turned off \citep{neyman1923application,rubin1974,lewis1974causation}. These two simulations are run in parallel (with and without the monosynapse), with identical (``frozen'') realizations of the noise processes in each condition. $\theta$ is thus defined as the difference in the height of the CCG peak between these frozen noise simulations. While focusing on the peak lag restricts somewhat the domain of action on which the synapse operates, it is also arguably the domain most deconfounded from background network activity (see section \ref{fig5unpack} in the Discussion). To normalize these comparisons, we study, in a given simulation, the quantity $|\theta - \hat{\theta}|/(|\textbf{R}|\cdot 10^{-3})$, which measures the bias in Hz, and {which we interpret as an estimate of $E[|\theta - \hat{\theta}|/(|\textbf{R}|\cdot 10^{-3})]$.} (There is an identification of [phase] space and time averages in this interpretation.) We call this the bias in the \textit{conditional synaptic synchrony rate} (or \textit{synaptic synchrony bias}, for short).
{(Recall that $|A|$ denotes cardinality when $A$ is a set, and absolute value otherwise.)}

Figure~{\ref{fig:monosynapse-estimation}}A measures the synaptic synchrony bias as a function of peak synaptic conductance ($g_0$) and synaptic decay time ($\tau_s$). Each pixel of each grid is estimated from a simulation of $10^8$ ms (this was considerably higher than was required for convergence). Here, the synaptic synchrony bias is negligible when $g_0$ is small or $\tau_s$ is small. In the small $g_0$ regime, estimation bias is small in part because $\theta$ itself is small. However, from the experiments of Figure~{\ref{fig:injected-synchrony}}, which represent the limiting case as $\tau_s \rightarrow 0$, we learned that the estimation bias is numerically undetectable regardless of $\theta$. Consistent with those observations, we observe the estimation bias tends toward 0 as $\tau_s \rightarrow 0$,  regardless of $g_0$, in Figure~{\ref{fig:monosynapse-estimation}}A. Hence, this suggests that the injected synchrony model can be broken by large $\tau_s$ (a natural consequence of the theory that will be discussed further in the Discussion). The estimation procedure works better for the numerically-optimized biophysical parameters than those of Table 1 (compare bottom and top panels), however, for the reasons above, the distinction between these two sets of parameters is not as important in the small $\tau_s$ regime.

In Figure~{\ref{fig:monosynapse-estimation}}B, we examine the estimation bias of $\hat{\theta}$ as a function of the postsynaptic gaussian noise ($\sigma_I$) and its timescale ($\tau_I$). In the insets of Figure~{\ref{fig:monosynapse-estimation}} we plot the firing rate, coefficient of variation, and CCG half-width averaged across the rows and columns of the grids. For a CCG, $C(\tau)$, CCG half-width is defined as $\sum_{\tau}\mathbbm{1}\{max(C(\tau))/2 > C(\tau)\}$. This was used instead of \textit{CCG Sharpness} (see Figure~{\ref{fig:monosynapse}} and Methods), because it was more computationally efficient in these large simulations and we only wish to note the average direction of change caused by each parameter here. 

The synaptic synchrony bias does not visibly depend on $\sigma_I$ and $\tau_I$ in Figure~{\ref{fig:monosynapse-estimation}}B because $|\textbf{R}|$ is used as the normalization term in this bias measure. To clarify this, we additionally plot the CCG half-width and $\theta$ corresponding with each pixel in the synaptic synchrony bias grids for the optimized parameter set. If we instead normalize the synaptic synchrony bias by $\theta$ instead of $|\textbf{R}|$ (i..e., $|\theta - \hat{\theta}|/\theta$), it is easier to see that $\theta$ is negatively correlated with the estimation bias (unlike in panels A and C, where it is positively correlated). Therefore, as in Figure~{\ref{fig:monosynapse-estimation}}A, we find the estimation bias chiefly depends on the sharpness of the CCG, consistent with the modeling assumptions. This additional normalization was required here because the bias grows as $\theta$ decreases, and because the magnitude of change effected by $\sigma_I$ is smaller than in the other panels.

Figure~{\ref{fig:monosynapse-estimation}}C demonstrates the influence of coincidence detection on the synaptic synchrony bias (via the postsynaptic adaptive threshold parameters, ($\alpha$ and $\tau_T$). We see that the coincidence detection property of the interneuron enhances the accuracy of the estimation. Recall that, in Figure~{\ref{fig:monosynapse-estimation}}C, $g_0$ and $\tau_s$ are fixed at the values of the asterisks in Figure~{\ref{fig:monosynapse-estimation}}A. This suggests that while the statistical model is broken by large $\tau_s$, to some degree the model's efficacy can be recovered in neurons with strong coincidence detection (see Discussion for further comments).

The trends we found throughout Figure~{\ref{fig:monosynapse-estimation}} were robust to various normalizations of the synaptic synchrony bias (by $S, |\vect{T}|$, and $\theta$), suggesting they do not result from minor distinctions in the definition of bias. Additionally, we also analyzed analogous grids for the naive estimator, and found that the synaptic synchrony bias was always higher using the naive estimator (numerator of Eq. \ref{eq:theta-closed-form}). The direction of bias was always of underestimation, as in Figure~{\ref{fig:injected-synchrony}}. 

Intuitively, one would expect there to be a relationship between the causal effect of a synapse, as formulated here (i.e., $\theta$), and generic measures of synaptic efficacy, and for this relationship to be particularly valuable in the regimes where $\hat{\theta}$ accurately tracks $\theta$. In the supporting information, we demonstrate this intuition numerically. In particular, we find a tight linear correlation between $\hat{\theta}$ and postsynaptic conductance, and illustrate its utility, in particular, when background network activity is a strong confound (see section S5).

\subsubsection{Parametric methods}
\label{sec:bootstrap-vs-jitter}

Our method for detecting and quantifying connection strength is motivated by the point of view of nonparametric and (in the case of the injection model) semiparametric statistics. This method can be computationally expensive when it is necessary to generate a large number of surrogate trains (as in $p$-value computation), in addition to possibly introducing technical concepts and methods that are unfamiliar. A baseline alternative, implicit in much of the neurophysiology literature, is given by parametric statistics based on classical distributions (e.g., the spike train is generated by an inhomogeneous Poisson process), and in particular by parametric bootstrap methods~\citep{Amarasingham2012, Kass2014}. A benefit of bootstrap in this setting can be that the null distribution of the test statistic can often be explicitly computed or at least quickly approximated, thanks to an explicit parametric model for the data (e.g., Poisson spiking). Nevertheless, we will demonstrate here that, even in approximately Poisson spiking regimes, and essentially because of the timescales involved, the  bootstrap test is a less sensitive synapse detector. Speaking to the broader issue of robustness to model specification, which we view as more fundamental, we provide a mathematical sketch in the Appendix demonstrating that the null hypothesis of conditional uniformity can essentially only be tested with a jitter method (section S4).

\begin{figure}
\includegraphics[width=\linewidth]{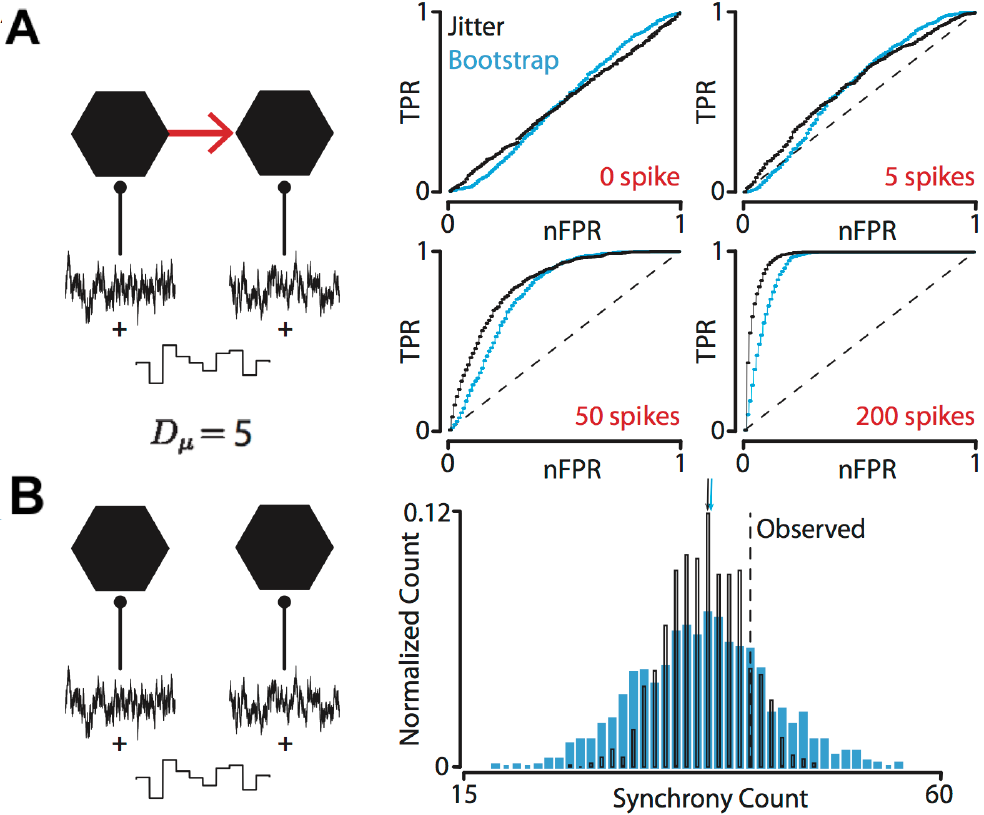} 
\caption{{\bf Bootstrap vs. jitter synapse detector.} \textbf{A:} (Left) A circuit composed of two monosynaptically connected standard LIF neurons was simulated over many trials ($1,000$ trials of $500$ s duration). The neurons were driven by both two independent white noise inputs and a common piecewise constant random input. The white noise parameters were chosen so that their firing was almost Poisson (mean: $V_T -10 \textrm{mV}$ with $V_T$ the threshold; standard-deviation: $5 \textrm{mV}$). The piecewise constant input was characterized by an interval length of $5 \textrm{ms}$ and an amplitude of $10 \textrm{mV}$ (i.e., $\textrm{maximum} - \textrm{minimum}$), making the Poisson firing highly inhomogeneous. The basic synapse model with spike injection was implemented. (Right) Modified ROC curves were computed based on the produced spike trains (nFPR: nominal false positive rate; TPR: true positive rate), for various amounts of injected synchrony (red). The $p$-values were computed using both jitter (black) and bootstrap (blue). The dashed line indicates the first diagonal. Note that all the jitter ROC curves lie above the bootstrap ones, confirming that the jitter has greater sensitivity of synapse detection. \textbf{B:} (Left) The connecting synapse was deleted from the circuit examined in A. (Right) For one simulation trial, a single synchrony count was observed (dashed line). Multiple surrogates ($\approx 1,000$) were generated from the original target spike trains, both using a jitter (black edge) and a bootstrap (blue) resampling methods. The resulting synchrony empirical distributions are shown. Note the similarity of the means (indicated by the arrows), but the larger spread for the bootstrap.}   
\label{fig:bootstrap-vs-jitter}
\end{figure}

We simulated the biophysical model of firing co-modulation in the absence of synaptic connection. Here we tuned the parameters such that the firing of each LIF neuron is approximately Poisson (see Methods)~\citep{Stevens1996}. More precisely, the spike trains are approximately inhomogeneous Poisson processes with an intensity function that follows the dynamics of the piecewise constant common input. We computed surrogate synchronies by generating numerous surrogate target trains ($100$ surrogates), and tabulating the amount of synchrony between the reference and each surrogate target train. In the jitter case, a surrogate train was generated by $\Delta$-jittering the original target spikes, with $\Delta=D_{\mu}$ (the piecewise constant input's interval length). In the bootstrap case, a surrogate train was generated by first measuring the firing rate in each $\Delta$-interval, then generating spikes from a Poisson process whose intensity function is given by the measured firing rate (constant in each $\Delta$-interval, but that can vary between intervals.) (Note that the Poisson bootstrap can be computed analytically, although we do not do that here.) In a first numerical experiment, we generated numerous spike trains ($1,000$) of the same duration ($500$ s). In half of the trains, we injected the same number of spikes in each target train following our basic synaptic model ($0$, $5$, $50$, or $200$ spikes). The fixed number of injected spikes ($\theta$) was then varied across iterations of the experiment. (From a neurophysiological perspective, the simulated data can be seen as originating from a cell population of independent neuron pairs.) We declare that a synaptic connection is detected whenever the $p$-value is lower or equal than a given threshold. To compare the sensitivity of the two methods, we computed the $p$-values for both methods for a wide range of threshold values, and estimated for each threshold value (i.e., nominal false positive rate) the probability of synapse detection (i.e., true positive rate), by taking the ratio of detected synapses to total connected pairs. The result of our experiment is shown in Figure~{\ref{fig:bootstrap-vs-jitter}A}. The verdict of the ROC is clear: here a jitter detector is more sensitive than a bootstrap, even in the Poisson regime in which the issue of model misspecification is presumed not to be an issue. This demonstrates the breakdown of bootstrap methods in highly nonstationary background firing regimes \citep{Amarasingham2015}, as we are reasoning that the background firing is approximately in the bootstrap null. (Note that the discrepancy is revealed by taking the total target spike count in a $5$ ms window centered around the CCG peak, as the synchrony definition ~\citep{Fujisawa2008, English2017}. See the Supplementary Appendix for a quantitative explanation of the necessity of this.)

The source of the difference in sensitivity can be intuited by looking at the distribution of surrogate synchronies for the two methods. Recall that the $p$-value can be computed as follows, $p = (1 + \sum_{j=0}^{N} \mathbbm{1} \big\{ S^{(j)} \geq S \big\})/ (N+1)$, where $N$ is the number of surrogates, $S$ is the total synchrony, and $S^{(j)}$ is a surrogate synchrony ($S^{(0)} = S$, by convention). We took one pair of trains at random from those simulated without a synapse. We computed numerous surrogate jitter and bootstrap synchronies ($1,000$ for each method), as described previously. One can observe that although both distributions are centered around the same average, the bootstrap synchrony distribution has a larger spread than the jitter synchrony one, as shown in Figure~{\ref{fig:bootstrap-vs-jitter}B}. Some calculations that clarify and explain these sensitivity effects mathematically are provided in the Appendix. 

\section{Discussion}
{The inference of microcircuit connectivity maps from \invivo recordings provides one of the important challenges in systems neuroscience. Such maps have been extracted in several studies~\citep{Fujisawa2008, Quilichini2010, English2017}, but rely heavily on statistical, sometimes heuristic, techniques. The focus has been on the strongest functional synaptic connections. It is natural to aim at more subtle interactions, in particular regarding the detailed interactions between learning and synaptic plasticity~\citep{Dupret2013, Miyawaki2016, Watson2016}, or the time course of weaker monosynaptic couplings, even in sparse firing conditions. With this goal of increased resolution in mind, we proposed a biophysical neuron model that reproduces two apparently paradoxical \invivo features: namely, the sharp cross-correlogram peak at short-latency, and high interspike interval variability. (These are paradoxical in the sense that while high variability of subthreshold noise may seem to underpin high interspike interval variability, in integrate-and-fire models, one may expect that low variability of subthreshold noise is associated with a sharp cross-correlogram peak.) We also developed rigorous statistical models for characterizing synaptic properties from fine-timescale spike pairwise correlations. We examined the physiological relevance of the statistical assumptions by applying it to spike data generated from the biophysical neuron model, and characterized biophysical regimes under which the statistical assumptions are valid, which gives insights in both directions between biophysics and statistics.

In parallel, from a more computational perspective, we illustrated a numerical mapping between a family of stochastic integrate-and-fire-type neurons driven by fast nonstationary noise and a semiparametric statistical model (the injected synchrony model) that accommodates the nonstationarities as nuisance parameters. An additional contribution is the formulation of monosynaptic characterization as the statistical estimation of the causal effect of the synapse on spike train correlations. Such a formulation may be relevant to more precise interpretation of, and calibration by, perturbation studies \citep{jazayeri2017}.

In the following, we discuss some alternative biophysical models. We propose physiological interpretations of the biophysical models. We point out some limitations of the injected synchrony model, and discuss how they might be resolved. }

\subsection{Physiological interpretation of the biophysical models}
Our synapse model is based on the one proposed and examined by~\citet{Ostojic2009}. Of note, the extreme sharpness of the CCG peak (Figure \ref{fig:juxta}), as found in physiological data, was not reproduced in their study. Furthermore, the authors cast doubt on the plausibility of this feature in the context of their model. We incorporated a few new components and tuned the parameters differently to recover these striking \invivo features. We detail here these modified elements along with their physiological meaning.

\subsubsection{Low variability of fast noise}
In our biophysical models, the ultra-precision of monosynaptic spike transmission relies critically on low variability of the fast input noise ($\sigma \ll 10$mV). This hypothesis was first proposed and theoretically motivated by~\citet{Herrmann2001} in the context of motoneuron experiments~\citep{Herrmann2002}. More precisely, the authors showed analytically that, in the low noise regime, the shape of the CCG peak is dominated by the positive part of the postsynaptic potential's time-derivative, $d \textrm{PSP}^{+}/dt$, making it extremely sharp ($d\textrm{PSP}/dt$ is only positive in the PSP's time-to-rise part). This fast noise input is intended to capture the collective result of numerous weak synaptic dynamics driven by uncoordinated presynaptic spike trains. Here low noise means that, over a short time window, the pool of asynchronous presynaptic cells fire sparsely (see~\citet{Gerstner2014} for detailed explanations). 

\subsubsection{Fast noise's large filtering timescale}
In contrast to~\citet{Ostojic2009}, we described the fast noise as colored (and not white). This choice was motivated by the result that, in this noise regime, spiking neurons are able to detect subtle input signal changes on the microsecond timescale, despite a typical millisecond-order membrane time constant (i.e., high-frequency encoding; see~\citet{Gerstner2014, Ostojic2015, Tchumatchenko2011} for extensive review of the theory, and experimental support). More precisely, the LIF neuron's spiking response to a deterministic transient input becomes instantaneous if the timescale of the colored noise is large enough compared to the membrane's time constant~\citep{Brunel2001}. In our model, this implies that even in the high noise regime ($\sigma \sim 10$mV), the PSTH (see supplemental  Figure~\ref{fig:ccgmech}A) would reflect the shape of the postsynaptic current (PSC) and not of the PSP, making it sharper regardless of the noise variability. Additionally, such a property makes possible the generation of nonstationary and co-modulatory firing with LIF neurons by simply controlling the input current, since the cell firing response follows this current instantaneously. From a physiological standpoint, the ``color'' of the noise originates from the intrinsic kinetics of the input synapses~\citep{Fourcaud2002, Gerstner2014}. The timescale of the noise is determined by the decay time constants of the synaptic input currents (i.e., PSC's half-width) -- a $10$ ms timescale can reflect the predominance of GABA\textsubscript{A} receptors. 

\subsubsection{Slow noise}
The slow input noise processes are an important distinguishing characteristic of the biophysical models we have studied here. These processes model synaptic inputs apart from the monosynaptic connection of interest, and are intended to exemplify synchronous network rhythms, such as gamma oscillations, as well as (unobservable) nonstationary background processes. This is the `background activity' in which the modeled cells are embedded~\citep{Buzsaki2006} and which acts as a source of coupling. A principal modeling assumption, underlying the plausibility of monosynaptic inference in nonstationary conditions, is that the dominant frequency (more generally, timescale) of such background processes is much slower than the synaptic spike transmission rate ($\sim 200$Hz)-- this is the so-called timescale separation hypothesis~\citep{Fujisawa2008}.

The choice of the slow (piecewise constant) input function is an expression of these ideas, and relates pedagogically to the null hypothesis of the jitter test. This null hypothesis is that, conditioned on the spike {\it counts} in the ($\Delta-$length) intervals, all placements of spikes are equally likely. The hypothesis is one of conditional uniformity, and it is the interval length that specifies the timescale of the process. The generality of this hypothesis stems from the fact that no assumptions are placed on the distribution of spike counts (in the intervals) themselves, which can be generated by processes of arbitrary complexity \citep{Amarasingham2012}. Thus the piecewise constant input function is intended as a dynamical source of the statistical (conditional) uniformity. The stochastic mechanism that generates the constants (the $\mu_k$'s in Equation \ref{piecewise-constant})  is simply a demonstration of arbitrary complexity. Here, we used such discontinuous inputs largely for mathematical and pedagogical simplicity. {To show the robustness of our principal conclusions, we report on experiments with continuous stochastic processes and other variations in Section \ref{background-timescale}.}

\subsubsection{Fast adaptive spike threshold}
As opposed to~\citet{Ostojic2009}, the threshold of our LIF model is dynamic and follows the membrane potential on a fast timescale~\citep{Platkiewicz2011, Fontaine2014, Mensi2016}. As shown by these last authors, the threshold's fast adaptation affects synaptic integration by reducing the effective timescale of somatic integration, enhancing the cell's coincidence detection property. More precisely, a LIF neuron with dynamic threshold can be seen as equivalent to a LIF neuron with fixed threshold but sharper PSPs~\citep{Platkiewicz2011, Mensi2016} -- their sharpness scales with the threshold time constant. Several authors argued that sodium channel inactivation is the main factor determining such threshold adaptation \invivo~\citep{Azouz2003, Henze2001, Wilent2005, Platkiewicz2011}, which received experimental support both \invitro and \invivo~\citep{Fontaine2014, Mensi2016}. Under this hypothesis, the timescale of threshold dynamics is primarily set by the inactivation time constant around voltage at spike threshold. But more broadly, it is reasonable to interpret fast threshold dynamics as a phenomenological model encompassing various mechanisms that increase the temporal precision of synaptic integration. A standard LIF model, as used by Ostojic et al., can account for a narrow peak in the CCG (see Figure 4 in~\citet{Ostojic2009}) but will not account for the sharp millisecond peak observed \invivo (see Figure 1E in~\citet{English2017} or Figure~{\ref{fig:monosynapse}} here). An additional coincidence detection mechanism was required for obtaining such a precise spike-spike transmission via a realistic synapse. These fast threshold dynamics are one of the simplest mathematical models for implementing an additional coincidence detection component, and would primarily mediated by a robust biophysical mechanism such as fast sodium channel inactivation at the site of action potential initiation.   

{\color{black} 
\subsection{The biophysical requirements of the injected synchrony model}
\label{fig5unpack}
Recall (Section \ref{sec:IS-Interpretation}) that the injected synchrony model in essence assumes each spike can be unambiguously (exclusively) labeled as caused by {either} a slow, background process, {or} a fast, synchrony-inducing process (e.g., the monosynapse). Accordingly, as verified in Figure~{\ref{fig:monosynapse-estimation}}, successful estimation of the causal effect of the monosynapse occurs in biophysical regimes where monosynaptic transmission is precise, as indicated by a thin CCG peak. Here,  \textit{precision} refers to how concentrated the relative timing of postsynaptic spikes caused by the monosynapse tend to be. Figure~{\ref{fig:monosynapse-estimation}} suggests the extent of monosynaptic precision determines the accuracy of our estimator, and hence the estimator serves as a good approximation in situations that slightly depart from the idealization of the {injected synchrony model} (as formulated in section \ref{injectedformulation}). Recall that, in Figure~{\ref{fig:monosynapse-estimation}}, $\theta$ was defined as the number of spikes caused by the monosynapse in the millisecond bin at the CCG peak. In future work, at minimum it will make more sense to model the spike trains on the continuum, and the interval of the presumed monosynaptic action will be specified independently of discretization. Similarly, it will also be more natural to weight the monosynaptic effect across temporal lags. The basic tradeoff here is the accumulation of power from capturing more of the monosynaptic interaction against increasing the confound of the background. Optimizing this tradeoff, ideally with perturbation data, is sensible and would presumably expand the regimes of high estimation accuracy which are plotted in Figure~{\ref{fig:monosynapse-estimation}}.

This tradeoff will also be set by the intrinsic biophysics of the neural circuit under study. Figure~{\ref{fig:monosynapse-estimation}A} showed that our estimator was highly accurate when $\tau_s$ is small. It is likely that the model breaks down when the PSP has a long timescale (high $\tau_s$ regime) because the timing of postsynaptic spikes caused by the monosynapses are likewise spread over a longer temporal interval. Such imprecise monosynaptically-induced spikes will not be absorbed into the synchrony count. Adapting the theorem in Section \ref{sec:injected-synchrony-model} correspondingly, this implies $E[\hat{\theta}] \leq \theta$ (under LEL or homogeneity). Indeed, in all of the simulations described in Figure~{\ref{fig:monosynapse-estimation}}, $\theta$ was consistently underestimated. This issue might be better addressed by incorporating temporal dynamics or multiscale properties into the statistical model~\citep{Harrison2014}, as discussed previously. The same reasoning likely underlies Figure~{\ref{fig:monosynapse-estimation}B and C}, where we found that the accuracy of our estimate increases with biophysical parameters that make monosynaptic spike transmission more precise.

Another important characteristic of our study is the we optimized the integrate-and-fire neuron parameters to fit the ACGs and CCGs in the \citet{English2017} data set. Two sets of parameters were used (Figure~{\ref{fig:biophysical-models}}), one optimized with loose constraints and one more conservatively informed by experimental measurements. The qualitative conclusions for each parameter set were the same, and the result was stronger when parameters were optimized to fit the \invivo data. It was precise monosynaptic transmission that characterized their data set.

A final point is that the parameters in Figure~{\ref{fig:monosynapse-estimation}} are varied separately for each plot (other parameters that are not being varied in a single plot are fixed). Therefore, while the asterisks denote the parameters as they are fixed, each plot in Figure~{\ref{fig:monosynapse-estimation}} suggests various distinct ways that accuracy can be recovered. Our choice of coincidence detection parameters in Table 1 are, in one sense, rather conservative (with, for the interneuron, $\alpha=0.75$ and $\tau_T=1$). While fitting the ACGs and CCGs, we found that we could fit both (1) the interneuron ACG and (2) the CCG Sharpness in the \citet{English2017} dataset even more precisely by making $\tau_T<1ms$ (e.g., $\tau_T\approx0.2$ms). These choices were used for the optimized parameters explored in Figure~{\ref{fig:monosynapse-estimation}} (bottom panels). 

}

\subsection{Alternative mechanisms for ultra-precision}
Feed-forward inhibition is another mechanism that can explain temporal ultra-precision in pyramidal-to-pyramidal synaptic transmission~\citep{Buzsaki1984, Pouille2001, Isaacson2011}. Following an excitatory input, a strong and delayed inhibition can prevent the postsynaptic cell from firing a few milliseconds after the excitatory delay, resulting in a millisecond-sharp CCG peak. This mechanism does not contradict our model though, as our assumptions only concern the postsynaptic cell's background input and membrane properties. Adding feed-forward inhibition to our model would presumably make short-latency spike coordination more precise and more robust.         

Any biophysical mechanisms known to increase the coincidence detection property of a cell should reinforce the temporal precision of synaptic spike transfer. For example, large average background synaptic conductances~\citep{Destexhe2003} or electrotonic spread between the dendrites and soma~\citep{Rall1960} can shorten the effective somatic integration timescale, and thus make short-latency spike pairwise coordination tighter. 

There are other biophysical mechanisms that could give rise to a cell's high-frequency encoding property, and therefore a sharp CCG. \citet{Richardson2010} and ~\citet{Droste2017} showed theoretically that non-Gaussian noise (i.e., synchronous state with large excitatory jumps as observed in~\citealp{DeWeese2006, Tan2014}) can create an ultra-fast cell's response to an input signal current, regardless of the noise timescale. \citet{Eyal2014} and~\citet{Ostojic2015} highlighted a similar effect using white noise-driven two-compartment neuron modeling, intended to abstract the interplay between soma and dendrites \textit{in vivo}. In a more general framework where action potential initiation is explicitly modeled, such as the exponential integrate-and-fire neuron, these results about high-frequency encoding will be all the more pronounced to the degree that spike initiation is sharper~\citep{Fourcaud2003}.


\subsection{Jitter background timescale} \label{background-timescale}
We showed in this study that the excess synchrony originating from a monosynaptic connection can be accurately estimated in some arguably realistic conditions, provided that the underlying slow background input timescale is known. But in practice we have, at best, a coarse understanding of this timescale. To better characterize the effect of the assumed timescale on estimation results, we first replicated the experiment and analysis underlying Figure~{\ref{fig:injected-synchrony}A} with various relaxations on the background process (as it is biophysically defined). First, rather than having a fixed $D_{\mu}$, we draw successive $D_{\mu}$'s independently from the uniform distribution on $(\Delta/2,3\Delta/2)$, where $E[D_{\mu}] = \Delta$. The accuracy of the estimates in Figure~{\ref{fig:injected-synchrony}A} is unchanged by this modification, when $\Delta$ is correct on average (Figure~{\ref{fig:relaxbackground}A}).

Likewise, we replicated Figure~{\ref{fig:injected-synchrony}A} but used instead a continuous slow background input, $\mu(t) = \mu_0 + \sigma_0 X(t)$, with $\mu_0 = -50$ mV, $\sigma_0 = 2.5$ mV, and $X$ a stochastic process described by the following dynamics,  
\begin{align} \label{eqn:smooth-input}
\begin{split}
D_{\mu} \frac{d X}{dt} &= -X + Y, \\
D_{\mu} \frac{dY}{dt} &= -Y + \sqrt{2 D_{\mu}} \xi,
\end{split}
\end{align}
where $Y$ is another stochastic process and $\xi$ is a Gaussian white noise process of zero mean and unit variance. These differential equations are arbitrary but incorporate an Ornstein-Uhlenbeck process that is commonly used for modeling synaptic noise. They were also designed so that the timescale $D_{\mu}$ is dominant while remaining simple enough. The accuracy of the estimates in Figure~{\ref{fig:injected-synchrony}A} are similarly unchanged by this modification (Figure~{\ref{fig:relaxbackground}B}). 

To better characterize the effect of the assumed timescale on estimation results, we replicated the experiment and analysis underlying Figure~{\ref{fig:injected-synchrony}A} while varying the jitter interval length ($D_{\mu}/4$, $D_{\mu}/2$, $D_{\mu}$, $D_{\mu}$, $2 D_{\mu}$, $4 D_{\mu}$, $10 D_{\mu}$, and the whole trial duration, $D_{\textrm{trial}}$). As shown in Figure~{\ref{fig:relaxbackground}C}, the closed-form estimate matches the true count as long as the interval length ($\Delta$) is less than or equal to the slow background input timescale, $D_{\mu}$. If the jitter length exceeds this timescale, the estimate agrees with the qualitative trend of the true value, but is clearly biased. Interestingly, in this experiment, jittering at $0.1$ or at $10,000$ seconds does not result in a significantly different estimate. This experiment suggests that the smaller the value of the jitter interval length, $\Delta$, the more confident we will be that the estimate is accurate, so long as the it is not so small as to interfere with the validity of the separation of timescale assumption.

\begin{figure}
\includegraphics[width=\linewidth]{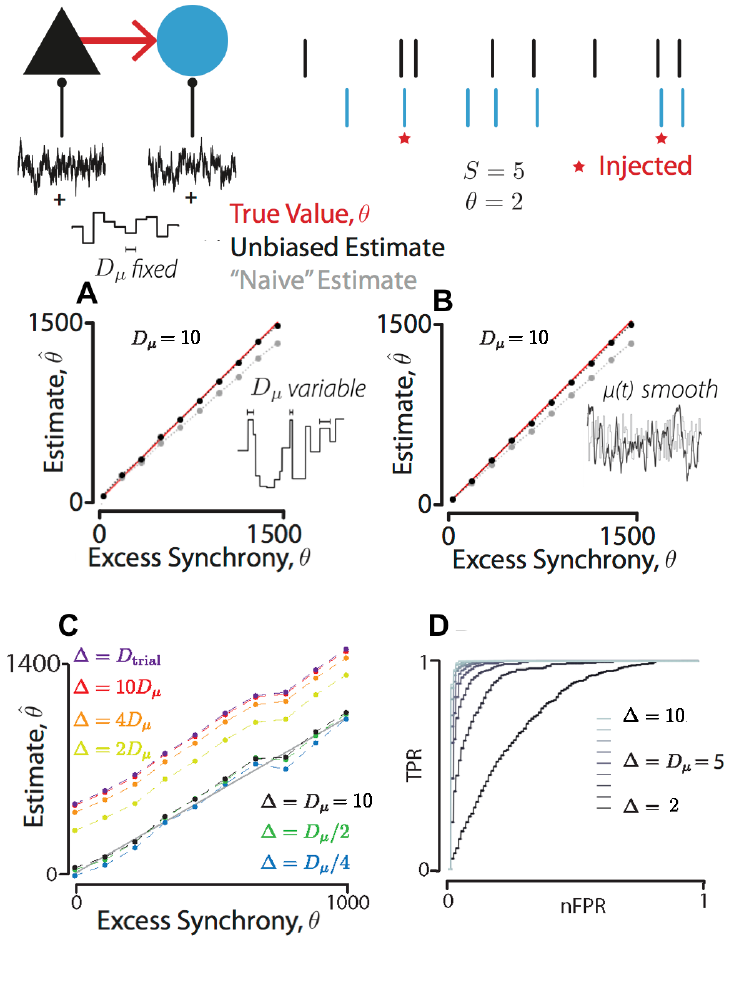} 
\caption{{{\bf Relaxing assumptions on the background input and its inference.} \textbf{A:} In previous figures, the idealized notion of a piecewise constant background input was used. Here, the experiment of Figure~{\ref{fig:monosynapse}}B is reproduced with the modification $D_{\mu} \sim U(\Delta/2,3\Delta/2)$ such that $E[D_{\mu}] = \Delta$. \textbf{B:} The same result is shown where the common input was generated using the system of equations~\ref{eqn:smooth-input}, a smooth stochastic process with timescale $D_{\mu}$. \textbf{C:}  The effect of choosing $\Delta$ on estimation for $\Delta \in \{D_{\mu}/4,D_{\mu}/2,...,D_{trial}\}$. The different partitions are indicated with different colors. Here, the piecewise
constant input’s interval length is $D_{\mu} = 10ms$.  \textbf{D:} The The effect of choosing $\Delta$ on detection, using precisely the same procedure as in  Figure~{\ref{fig:bootstrap-vs-jitter}}. Here, $\Delta \in \{1,2,...,10\}$ and $D_{\mu} = 5 ms$. Detection only breaks down when the separation of timescales assumption is violated.}}   
\label{fig:relaxbackground}
\end{figure}

Regarding testing, the statistical power of the jitter test will also depend on $\Delta$. To illustrate this, we replicated the experiment and analysis of the previous section (`Parametric Methods'), fixing the number of injected spikes at $50$ and varying the jitter interval lengths ($\Delta \in \{2, 3,...,10\}$) with $D_{\mu} = 5$. The result of this experiment is shown in Figure~{\ref{fig:relaxbackground}D}. We observe that the greater the jitter timescale, the greater the power of the jitter test. This is again intuitive; comparing among nested null hypotheses, increasing $\Delta$ further restricts the null hypothesis \citep{Harrison2014}. In contrast, detection worsens appreciably only when $\Delta$ approaches the synaptic timescale. The combination of the last two numerical experiments confirms that there is an optimal jitter timescale that controls the trade-off between estimation bias and the test power: $\Delta$ should be low enough so that the synaptic strength is well estimated for any correctly detected synapse, but large enough to control the false detection rate.} 

In practice, investigators will find it useful to examine how neurophysiological conclusions of interest vary with $\Delta, $
which corresponds to the common statistical technique of evaluating robustness by comparing statistical conclusions across variations in statistical assumptions. Figure 3 is an example of such an approach (cf., \citealp{Harrison2014} and Figure 5 of \citealp{martin2015spike}). Developing a principled framework for such interpretations may be a fruitful goal for future research, and may be related to the goal of developing a multiscale analogue of the injected synchrony model.

{
\subsection{Alternative methods for inferring synaptic connectivity}

{The central feature of the approach to synaptic characterization described here is the identification of monosynaptic interactions with fine-timescale interactions using nonparametric techniques in a principled way. There are two primary reasons for taking this approach: i) the observation of finely-timed interactions in association with monosynaptic connections \invivoe, and ii) a desire to be agnostic with respect to the background model (the input to a postsynaptic neuron that does not derive from the monosynapse). The approach is similar in spirit to that of \citet{Fujisawa2008}, differing in particular features. Among these differences are principled approaches to formulating the null model of no interaction (interval jitter rather than spike-centered jitter;  \citealp{Platkiewicz2017}) and to measuring strength of effect (i.e., deviation from the null hypothesis). A consequential choice was to use coupled background noise models that operate just above the timescale of monosynaptic interaction. Our motivation here is to be as agnostic as possible about the background model. Using a background model that, in an intuitive sense, closely mimics monosynaptic interaction is akin to the classical technique of optimizing over a null hypothesis to construct a hypothesis test. This should act to minimize false positives, provided that the separation of timescale hypothesis is valid. Another presumption, here of convenience, is that the action of the monosynaptic effect is concentrated on the lag at which the CCG peaks, as discussed in section \ref{fig5unpack}. Finally, focusing on the synaptic detection (hypothesis testing) component exclusively, here we do not adopt the conservative stance of correcting for multiple tests across CCG lags as in \citep{Fujisawa2008}. 
Regarding the broader motivation for the general strategy of inference, the use of jitter-derived tests and their analogues are then intimately tied to the choice of using the conditionally uniform null hypothesis to formulate separation of timescale: in a certain sense, robustness then necessitates a jitter-type test (see Section S4 in the Appendix).

    Some alternative methods of note make finer-grained modeling choices \citep{Aertsen1985, Kobayashi2019}. Of recent relevance is the study of~\citet{Kobayashi2019}, where the authors model the cross-correlogram using a generalized linear model, with a Bayesian prior on parameters, including on the covariance of the background contribution to the firing rate and then compute a MAP estimate of the parameters numerically. The key issue is one of interpretation, in evaluating the suitability of the statistical assumptions (for example, the form and the role of the prior, and the role of model selection penalties in model-fitting and its suitability for the monosynapse problem). The difficulty of such an assessment is a reflection of the challenges of the monosynapse problem.  Valuably, Kobayashi et al. validate these judgements using biophysical models involving networks of randomly-connected neurons. Comparing differing approaches such as these will be a long term challenge. An aspect to immediately isolate is the background coupling the Kobayashi model induces and upon which monosynaptic dynamics are imposed. One may expect that randomly-connected networks with uncoupled external inputs are not necessarily designed to optimize the high degrees of anatomical or functional common input which would generate fast comodulatory drive to a given neuronal pair,  as discussed above. Another key aspect to compare is the timescale of fast monosynaptic dynamics, which is the focus of this work. 
    
    Kobayashi et al. show that, using their biophysical model, their proposed inference methods outperform the Fujisawa et al. method in the sense of false negatives (e.g., missed existing connections), in particular for inhibitory connections. The false positive rate, for those particular implementations, is comparable. Thus one immediately interesting and important question this raises is how these and other performance differences depend on variations in the anatomical assumptions in the network model, particularly the statistical distribution of common anatomical input, across neuron pairs.
    
    
    
    
  }

}

\subsection{Toward a theory of single trial data analysis}

Systems neuroscientists are concerned that in typical data-analytical problems conventional approaches to aggregating statistical information across trials, such as averaging, do not apply~\citep{Amarasingham2006,churchland2007techniques,Gao2015}. A more detailed look at this concern reveals the fundamental issue is often overly simplified, even ill-defined, noise models~\citep{Amarasingham2015}. The Gaussian noise-driven LIF neuron is a case in point. The application of conditional modeling to timescale specification was designed to address such issues. However, this framework can abstract away the biophysical foundations of spike train models, confounding properties that are typically understood in biophysical terms ~\citep{Gao2015}. 

We can anticipate a hierarchy of models, at varying levels and blends of physiological and statistical detail. Towards this development, the single- and two-timescale models here could be extended to account for the multiple timescales characterizing background network activity, as emphasized by~\citet{Buzsaki2006}. The biophysical models presented here, and   numerous biologically-plausible variations~\citep{Gerstner2014}, might be incorporated into the statistical analysis of large-scale extracellular spike data, to further constrain the inference of microcircuit connectivity diagrams. Such a combination may address the issue of overly constrained noise models, and other challenges. As one potentially-fruitful example in this vein, it has been suggested that
population spike activity cannot be summarized by independent pairwise spike cross-correlations~\citep{Schneidman2006, Pillow2008, Gerhard2013}, a mismatch proposed to arise from local connectivity patterns (e.g., triplet connectivity~\citep{Song2005, Jiang2015, Cossell2015}).

\section{Methods}
\subsection{Biophysical modeling}
\subsubsection{Single neuron}
The firing dynamics of a cortical neuron was simulated using a leaky integrate-and-fire model with a fast adaptive threshold -- an inactivating leaky integrate-and-fire neuron (iLIF)~\citep{Platkiewicz2011, Mensi2016}. A spike is emitted whenever the neuron's instantaneous membrane potential $V(t)$ crosses an instantaneous threshold $V_T(t)$. Immediately after spike triggering, $V$ is reset to a fixed voltage $V_R$ and can be held to this value for a few milliseconds (i.e., refractory period, which was not considered in our simulations). Between two spikes, the dynamics of $V$ and $V_T$ are linear,  
\begin{eqnarray}
\tau_T \frac{dV_T}{dt} &= -V_T + f(V),\\
\tau_m \frac{dV}{dt} &= -V + I,
\label{eq:iLIF}
\end{eqnarray}
where the voltage $I$ is the neuron's input; the function $f(V) = V_{T0} +  \alpha \, \textrm{max}(0, V-V_i)$ is rectified linear, with $V_{T0}$ and $V_i$ the baseline threshold and the inactivation voltage, respectively; the parameters $\tau_m$ and $\tau_T$ are respectively the passive membrane time constant and the threshold time constant. The dynamics of $I$ was described using an Ornstein-Uhlenbeck process,
\begin{equation}
\tau_I \frac{dI(t)}{dt} = -I(t) + \mu(t) + \sigma_I \sqrt{2 \tau_I} \xi(t),
\label{eq:OU}
\end{equation}
where the stochastic term $\xi$ is a Gaussian white noise process of zero mean and unit variance. For two times $t$ and $t'$, we have $\E[\xi(t) \xi(t')] = \delta(t'-t)$, with $\delta$ the Dirac Delta function. The parameters $\tau_I$ and $\sigma$ are respectively the colored noise's time constant and standard-deviation. To incorporate a nonstationary drive, the time-dependent variable $\mu(t)$ was made stochastic. It was modeled as a piecewise constant function on equal-length intervals of size $D_{\mu}$, \newline $\mu(t) = \displaystyle \sum_{i=1}^{m} \mu_i \mathbbm{1} \big\{ (i-1) D_{\mu} \leq t < i D_{\mu} \big\}$, where $m$ is the number of intervals of size $D_{\mu}$ in the trial, and the values $(\mu_1, \ldots, \mu_m)$ were independently and uniformly sampled in $[\mu_{\textrm{min}},\mu_{\textrm{max}}]$. We have, as a result, \newline $\E[I(t)I(t')] = \mu_i \mu_j +\sigma^2 \exp \big( -|t'-t|/\tau_I \big)$, $t$ and $t'$ belonging respectively to the $i$'th and $j$'th intervals. It is important to note that the membrane potential, $V$, fluctuates around the mean of $I$, $(\mu_{\textrm{min}} + \mu_{\textrm{max}})/2$, whose value is typically between $V_R$ and $V_{T0}$. 

\subsubsection{Background co-modulation}
The model of firing co-modulation between two cortical neurons was a simple extension of the single neuron one. The equations of $V$ and $I$ for both neurons were the same as Eqs.~\ref{eq:iLIF} and~\ref{eq:OU}, except that here there are two independent white noise terms, $\xi_1$ and $\xi_2$, where the indices $1$ and $2$ denote the two neurons of the pair. For distinct times $t$ and $t'$ we therefore have $\E[\xi_1(t) \xi_2(t')] = 0$, but $\E[I_1(t)I_2(t')] = \E[\mu(t) \mu(t')]$ since the two neurons share the same exact input mean. In other words, the cross-correlation function of $I_1(t)$ and $I_2(t)$ is completely determined by the auto-correlation function of $\mu(t)$. 

\subsubsection{Monosynaptic transmission}
Monosynaptic coupling between two neurons was modeled as in~\citet{Ostojic2009} using an input conductance, $g_S$. The membrane equation becomes,
\begin{equation}
\tau_m \frac{dV}{dt} = -V + I + \frac{g_s}{g_l} (V-E_s),
\label{eq:Vm-monosynaptic}
\end{equation}
with $E_s$ the synaptic reversal potential -- if $V=E_s$, the synapse has no influence on $V$. The synaptic conductance is instantaneously incremented by an amount $g_0$ after the transmission of each presynaptic spike, characterized by a fixed delay, $\delta_s$. Between two spikes, its dynamics are linear,
\begin{equation}
\tau_s \frac{dg_s}{dt} = -g_s.
\label{eq:Vm-monosynaptic}
\end{equation}
Following a presynaptic spike emission at time $t_0$, we have $g_s(t) = g_0 \exp \big( -(t-t_0-\delta_s)/\tau_s \big) \mathbbm{1} \{t \geq t_0+\delta_s\}$.

\subsection{Numerical simulations}
All simulations of biophysical neuron models were performed using Brian, a spiking neural network simulator in Python~\citep{Goodman2008}. The spike cross-correlograms were computed using the corresponding function provided in phy, an open source neurophysiological data analysis package in Python~\citep{Rossant2016}. For the remaining simulations and generated data analysis, we used NumPy and SciPy (numeric and scientific modules of Python). All the code used to produce the main results will be available online (\url{https://github.com/aamarasingham}).

\subsubsection{Model Parameters}
Table 1 parameters are shown in this section. The optimized parameters, which were reported in the Results, are also shown in Table 2. As mentioned previously, Table 2 values were optimized with loose constraints. The values in Table 1, in contrast, were set to more parsimonious values by pulling from experimental measurements. Although note that many such values are typically measured \invitro (see Results), which is yet another motivation for using two sets of parameters.

\begin{table}[!htbp]
\centering
\caption{\label{tab:biophysical-parameters} {\bf Biophysical Parameters Values.}}
\begin{tabular}{|l|l|l|l|}
\hline
\multicolumn{1}{|l|}{\bf Parameters} & \multicolumn{1}{|l|}{\bf Pyramidal} & \multicolumn{1}{|l|}{\bf Interneuron} & \multicolumn{1}{|l|}{\bf Poisson}\\ \thickhline
$\tau_m$  (ms)  & $10$  & --     & -- \\
$V_R$     (mV)  & $-60$ & --     & -- \\
$V_i$     (mV)  & $-60$ & --     & -- \\
$\tau_I$  (ms)  & $10$  & --     & $\sim$ $0$ \\
$D_{\mu}$ (ms)  & $10$  & --     & $5$        \\
$\mu_{\textrm{max}}$ (mV) & $-45$ & -- & $-35$ \\
$\mu_{\textrm{min}}$ (mV) & $-55$ & -- & $-45$ \\
$\tau_s$  (ms)  & $\times$   & $3$     & $\times$ \\
$\delta_s$ (ms) & $\times$ & $1.5$ & -- \\
$E_s$  (mV)  & $\times$      & $0$     & $\times$ \\
$\tau_T$  (ms)  & $7$   & $1$    & $\sim \infty$ \\
$V_{T0}$  (mV)  & $-55$ & $-57$  & $-50$ \\
$\alpha$        & $1$   & $0.75$ & $0$   \\
$\sigma_I$ (mV) & $6$   & $2$    & $5$   \\\hline
\end{tabular}
\begin{flushleft}  `--' : same value as in the column to the left;
`$\times$' : not applicable.
\end{flushleft}
\end{table}

\begin{table}[!htbp]
\centering
\caption{\label{tab:opt-parameters} {\bf Optimized Biophysical Parameters Values.}}
\begin{tabular}{|l|l|l|l|}
\hline
\multicolumn{1}{|l|}{\bf Parameters} & \multicolumn{1}{|l|}{\bf Pyramidal} & \multicolumn{1}{|l|}{\bf Interneuron}\\ \thickhline
$\tau_m$  (ms)  & $20.06$  & $19.59$\\
$V_R$     (mV)  & $-60$ & --\\
$V_i$     (mV)  & $-52.80$ & $-59.27$\\
$\tau_I$  (ms)  & $11.76$  & $4.97$ \\
$D_{\mu}$ (ms)  & $10$  & --\\
$\mu_{\textrm{max}}$ (mV) & $-30.95$ & $-43.95$\\
$\mu_{\textrm{min}}$ (mV) & $-63.83$ & $-66.13$\\
$\tau_s$  (ms)  & $\times$   & $3$ \\
$\delta_s$ (ms) & $\times$ & $1.5$\\
$E_s$  (mV)  & $\times$      & $0$\\
$\tau_T$  (ms)  & $13.54$   & $0.22$\\
$V_{T0}$  (mV)  & $-44.96$ & $-57.7$\\
$\alpha$        & $1$   & $0.77$\\
$\sigma_I$ (mV) & $14.72$   & $2.93$\\\hline
\end{tabular}
\begin{flushleft}  `--' : same value as in the column to the left;
`$\times$' : not applicable.
\end{flushleft}
\end{table}

\subsubsection{Spike count computation} \label{sc-computation}
In all but Figure~{\ref{fig:bootstrap-vs-jitter}}, synchrony was computed by first partitioning time into bins of a certain width ($1$ ms). The spike trains were then binned (i.e., $1$ if any spikes in the bin, $0$ otherwise). If for a given bin, a spike was observed in each train of the pair, a synchrony is counted. Naturally, we previously shifted the postsynaptic train backwards in time by the number of bins corresponding to the implemented synaptic latency.    

In Figure~{\ref{fig:bootstrap-vs-jitter}}, we computed first the cross-correlogram between the reference and target trains. The synchrony count is then the sum of the raw spike count in each CCG time bin (of $0.1$ ms width) over a window centered around the CCG peak (of $5$ ms width). 

The ACG's and CCG's were normalized by dividing the spike count in each time bin by the number of reference spikes and the bin width. Under stationary conditions, the normalized count would represent an estimate of the target cell's instantaneous firing rate relative to a reference spike time.

\subsection{Semiparametric analysis of spike trains}

\subsubsection{Preliminary notation}
\label{ccg-def}
As described in the main text, a spike train, $\vect R$ for example, is modeled in discrete time, and we write $\vect  R=(R_1, \ldots, R_{|\vect R|} )$, where $R_i$ is the time of the $i'$th spike in $\vect R$, and $|\vect R|$ is the number of elements in $\vect R$. Time is partitioned into equal-length intervals of size $\Delta$, and the spike counts in $\vect R$ in the intervals are denoted $\vect  N(\vect R) = (N_1(\vect R), \ldots, N_m(\vect R))$ where $m$ is the number of intervals~\citep{Amarasingham2012}. 

Here trains $\vect R$ and $\vect T$ will be referred to as the reference and target, respectively. Given the spike trains $\vect R$ and $\vect T$, their synchrony count is defined as
\begin{equation}
S = \displaystyle  \sum_{i=1}^{|\vect T|} \sum_{j=1}^{|\vect R|} \ind \{T_i-R_j = \textrm{lag}\},
\label{syncheq}
\end{equation}
where $\textrm{lag} \geq 0$ represents the synaptic delay in units of time bins. The discretization specifies the resolution of synchrony. Throughout the paper, the cross-correlogram (CCG) is derived from Eq.~\eqref{syncheq} by letting \textit{lag} be variable and dividing the synchrony count for each \textit{lag} by $|\vect R|$. Similarly, auto-correlagrams (ACG) are the case where the reference and target train are the same.

In addition, we impose the condition that all intervals that contain a spike in $\vect T$ also contain a spike in $\vect R.$ (Such target spikes cannot contribute to the synchrony count by definition. They are not included in  $\vect T$).

Finally, it will be useful to define 
\begin{equation}
S_i =\sum_{j=1}^{|\vect R|} \ind \{T_i-R_j = \textrm{lag}\},
\end{equation}
which indicates whether the $j'th$ target spike is synchronous. Then $S=\sum_{i=1}^{|\vect T|} S_i.$

As mentioned above, in practice the postsynaptic spike train can be shifted in time, in advance of any analysis, so that the lag can effectively be treated as 0, and this is the practice we adopted in this study.



\subsubsection{Injected synchrony model} \label{sec:injected-synchrony-model}
When evaluating a cell pair that is putatively monosynaptic, the candidate presynaptic train is chosen as the reference spike train and the candidate postsynaptic one as the target. In the following, we will assume the reference train to be fixed and the target train to be random (i.e., we are reasoning conditionally on the reference). The injected synchrony model is that (conditioned on the reference train $\vect R$), the target train $\vect T$ is a superposition of two point processes: an ``injected'' train, $\vect I$, in which every spike is synchronous with a reference spike; a ``background'' train, $\vect B$, that is in the $\Delta$-jitter null. $\vect B$ is conditionally uniform, conditioned on $\vect N(\vect B)$ and $\vect R$: 
\begin{equation}
    P( \vect B=\vect b | \vect N(\vect B)=\vect N(\vect b),\vect R=\vect r ) = c( \vect N(\vect b), \vect r),
\end{equation}
for some function $c(\cdot,\cdot)$. (In other words, the conditional likelihood that $\vect B$ takes the value $\vect b$, given knowledge of $\vect N(\vect B)$ and $\vect R$, depends on $\vect b$ only through $\vect N( \vect b)$.) $\vect T = \vect B + \vect I$, where $+$ indicates superposition. Note that in this model two spikes can be in the same bin in the superposition. We think of the latter as an approximation. 

The estimation problem is to infer the number of injected synchronies, $\theta=|\vect I|$, where `$|.|$' indicates the number of elements in a set. Reasoning conditionally (or parametrically, in the case of $\theta$), we treat $\theta$ and $\vect R$ as fixed, and we will assume that the background timescale $\Delta$ is known.

We will use the following notation:
\begin{enumerate}
	\item Let $W_i$ identify the interval (with respect to the interval partition) in which target spike $T_i$ falls. We will collect these as $\bm W = (W_1,W_2,...,W_{|\bm T|})$. 
	\item Let $N_{r,i}$ denote the number of spikes in the reference train in window $W_i$. We will collect these as $\bm N_{ r}=(N_{r,1},N_{r,2},..\\..,N_{r,|\bm T|}).$ $\bm N_{r}$ is a function of $\bm R$ and $\bm W.$ (Note that, by virtue of $\bm T$'s construction, $N_{r,i}>0$ for all $i$.)
	\item Let $\ell_i$ take the value 1 if spike $T_i$ is an injected spike (derived from $\bm I$), let it take the value 0 if it is a background spike (derived from $\bm B$). We will collect these as $\mathcal L=(\ell_1, \ell_2, ..., \ell_{|\bm T|}).$ We have $\theta = |\bm I| = \sum_i \ell_i$.
	\item Denote $\bar{r}$ as $(1/|\vect T|)\sum_{i=1}^{|\vect T|} N_{r,i}$.
\end{enumerate}




\noindent {\bf Theorem.} {\it Assume that either of the following two conditions hold:
\begin{enumerate}
    \item   {\bf Condition 1 [$\bm R$ `Homogeneous'].} {\it $\bm R$ is `homogeneous,' in the sense that non-zero values of $N_{r,i}$ are constant (not to be confused with homogeneity in the sense of stationarity).\footnote{Since $\bm R$ is in fact observable, this is a potential observation rather than an assumption.} [The principal example is the sparse case where $N_{r,i}=1$ when non-zero (or approximately so), which will generally be more relevant for either low firing rates or, in particular for the monosynapse problem,  small $\Delta$.]
    \item {\bf Condition 2 [All labellings equally likely (LEL).]} {\it Conditioned on $(\bm W, \bm R),$ all $\mathcal L$ are equally likely (LEL), in the sense that
    \begin{equation}
    P( \mathcal L | \bm W, \bm R) = \binom{|\bm T|}{\theta}^{-1} \mathbbm{1} \{ |\mathcal L|=\theta\}.
   \end{equation} } }
\end{enumerate}
Then 
\begin{equation}
\hat{\theta} = \frac{ S - \frac{\bar{r}}{\Delta}|\bm T| }{1-\frac{\bar{r}}{\Delta}}
\end{equation}
is an unbiased estimator of $\theta.$} }

\noindent {\bf Proof.} 
Compute

\begin{align}
\,&\phantom{=}E[S|\bm W,  \bm R, \mathcal L] \nonumber \\
&= E \left. \left[ \sum_{i=1}^{|\vect T|} S_i \mathbbm{1} \{ \ell_i=1 \} + \sum_{i=1}^{|\vect T|} S_i \mathbbm{1} \{ \ell_i=0 \} \right| \bm W, \bm R, \mathcal L \right] \nonumber \\
&= E \left. \left[ \theta + \sum_{i=1}^{|\vect T|} S_i \mathbbm{1} \{ \ell_i=0 \} \right| \bm W, \bm R, \mathcal L \right] \nonumber \\
&= \theta + \sum_{i=1}^{|\vect T|} E \left. \left[  S_i \mathbbm{1} \{ \ell_i=0 \} \right| \bm W, \bm R, \mathcal L \right] \nonumber \\
&= \theta + \sum_{i=1}^{|\vect T|} \mathbbm{1} \{ \ell_i=0 \} \underbrace{E \left. \left[  S_i \right| \bm W, \bm R, \mathcal L \right]}_{\frac{N_{R,i}}{\Delta}} \nonumber \\
&= \theta + \sum_{i=1}^{|\bm T|} \frac{N_{R,i}}{\Delta} \mathbbm{1} \{ \ell_i=0 \}
\label{eq:hanproposition}
\end{align}

\noindent Then assume, first, that Condition 1 holds. By (\ref{eq:hanproposition}), in this case $E[ S | \bm W, \bm R, \mathcal L] = \theta + \frac{\bar{r}}{\Delta}( \bm |T| - \theta)$ independently of $\mathcal L$. Thus
\begin{equation}
E[ S | \bm W, \bm R] = \theta + \frac{\bar{r}}{\Delta}( |\bm T| - \theta).
\end{equation}
Rearranging, we have
\begin{equation}
\frac{ E[ S | \bm W, \bm R] - \frac{\bar{r}}{\Delta}|\bm T| }{1-\frac{\bar{r}}{\Delta}} = \theta,
\end{equation}
if $\Delta>1$. Taking expectations on both sides:
\begin{equation}
E \left[ \frac{ S - \frac{\bar{r}}{\Delta}|\bm T| }{1-\frac{\bar{r}}{\Delta}} \right] = \theta,
\end{equation}
where we have used the fact that $\bar{r}$ is determined by $(\bm W,\bm R),$ and the general fact that $f(Z)E[X|Z]=E[f(Z)X|Z],$ for any random variables $X,Z$ and function $f$.
So that (under homogeneity)
\begin{equation}
\hat{\theta} = \frac{ S - \frac{\bar{r}}{\Delta}|\bm T| }{1-\frac{\bar{r}}{\Delta}},
\end{equation}
defined if $\Delta>1$, is an unbiased estimator of $\theta$. (If $\Delta=1,$ there is no information in the spike trains about $\theta$.) 

Now consider the case that Condition 2 holds. Starting with Eq. (\ref{eq:hanproposition}), explicit calculation shows that 
\begin{equation}
   E[ S | \bm W, \bm R] = \theta + \frac{|\bm T|-\theta}{|\bm T|}\sum_{i=1}^{|\bm T|}\frac{N_{R,i}}{\Delta}.
\end{equation}

\noindent Then, using reasoning analogous to the previous condition, we once again obtain

\begin{equation}
\hat{\theta} = \frac{ S - \frac{\bar{r}}{\Delta}|\bm T| }{1-\frac{\bar{r}}{\Delta}}
\label{eq:theta-closed-form}
\end{equation}
as an unbiased estimator of $\theta. \blacksquare$

In the absence of homogeneity, and using a similar idea, estimators that are bounded below and above $\theta$ in expectation can be constructed by optimizing $E[ S | \bm W, \bm R, \mathcal L]$ over $(\theta,\mathcal L).$ We do not develop this approach here, but it is reflected in the confidence intervals developed next.

\subsection{Confidence intervals for $\theta$}
Here is a a general approach to constructing a confidence interval for $\theta$ following the lines above. The idea is to construct hypothesis tests for null hypotheses of the form $H_0: \theta=j,$ and then to invert those tests to form a confidence interval for $\theta.$

Compute 
\begin{align} \label{eq:c+}
c^+(\bm W, \bm R, j) &:= \\
& \argmin_k \left\{ k \left| \max_{ \{\mathcal L | |\mathcal L|=j\} } P( S \geq k | \bm W, \bm R, \mathcal L ) \leq \alpha/2 \right. \right\} \nonumber
\end{align}
and 
\begin{align}
c_-(\bm W, \bm R, j) &:= \label{eq:c-} \\
&\argmax_k \left\{ k \left| \max_{ \{\mathcal L | |\mathcal L|=j\} } P( S \leq k | \bm W, \bm R, \mathcal L ) \leq \alpha/2 \right. \right\} \nonumber
\end{align}
It follows that $\{ S \geq c^+(\bm W, \bm R, j) \} \cup \{ S \leq c_-(\bm W, \bm R, j)  \}$ is a critical region for an $\alpha$-level test of $H_0: \theta=j$. Then `invert' the tests \citep{casella2002statistical}: the set

\begin{equation} \label{eq:CI}
C(S,\bm W,\bm R,\alpha) = \{ j \, | \, c_-(\bm W, \bm R, j) \leq S \leq c^+(\bm W, \bm R, j) \}
\end{equation}
is a (contiguous) $(1-\alpha)$-level confidence interval for $\theta$, meaning that
\begin{equation}
P( \theta \in C(S,\bm W,\bm R,\alpha) ) \geq 1-\alpha.
\end{equation}
For any $(\bm W, \bm R, \mathcal L),$
\begin{equation}
S \left| \bm W, \bm R, \mathcal L \right. \sim |\mathcal L| + \sum_{i=1}^{|\vect T|} Z_i \ind\{ \ell_i=0  \}
\end{equation}
where $\sim$ signifies equality of (probability) distribution, $Z_i$ is a Bernoulli random variable with parameter $N_{R,i}/\Delta$, and $Z_1, Z_2, $ ..., $Z_{|\vect T|}$ are independent. Thus $P( S = k | \bm W, \bm R, \mathcal L )$ can be computed by convolution (see Section A.6 of \citet{amarasingham2011conditional}). The only issue that then remains is the computational feasibility of repeatedly evaluating (\ref{eq:c+}) and (\ref{eq:c-}) to compute (\ref{eq:CI}). It turns out that the confidence interval $C(S,\bm W,\bm R,\alpha)$ can be computed efficiently because, in brief, i) the optimizations over $\mathcal L$ in the above expressions are straightforward and independent of $k$, ii) $\max_{ \{ \mathcal L \, | \, |\mathcal L|=j \} }  P(S \geq k|\bm W, \bm R, \mathcal L)$ and $\max_{ \{ \mathcal L \, | \, |\mathcal L|=j \} }  P(S \leq k|\bm W, \bm R, \mathcal L)$ can be computed exactly by convolution, and iii) there is a way to rearrange the computations so that (\ref{eq:c+}) and (\ref{eq:c-}) do not have to be computed independently for each $j$.  The validity of the intervals do not require stronger assumptions of the model. 

It also turns out that, in analogy to the estimation problem and using similar ideas, shorter exact (and contiguous) confidence intervals can also be computed by invoking the LEL assumption. Thus, in addition, both intervals are simultaneously useful in the sense that their discrepancy can be used to measure the effect of the LEL and homogeneity assumptions. 

The finer details of these procedures are outside the scope of this study and will be reported elsewhere. Code for computing $C(S,\vect W, \vect R, \alpha)$ will be available at \url{https://github.com/aamarasingham}.

\subsection{Analysis of \textit{in vivo} data}
To reproduce \invivo firing in single integrate-and-fire neurons, we proceeded in three steps. First, we regularized the shape of the \invivo ACG's of both interneurons and pyramidal cells with a simple model, a composition of two exponentials, so that their features could be summarized in six dimensions. Second, we determined the most representative pyramidal cell and interneuron in the data (defined according to the k-medoid algorithm, as described below). Thirdly, we fit the integrate-and-fire neuron model to the ACG model of the medoid ACGs using global optimization methods. These methods served to help reproduce representative 
\invivo firing statistics in the integrate-and-fire simulations, and provided a neutral and reproducible approach for approximating some parameters that are not well-understood experimentally. The composition-of-exponentials model used to fit the \invivo ACGs is

\begin{align}
\begin{split}
M(\tau) = &(\beta_2 + \beta_4)\: \exp\left[\frac{(\tau-\beta_6)^2}{\beta_1}\right] \: \mathbbm{1}\{\tau \leq \beta_6  \} \\
 & + \left(\beta_2\:\exp\left[\frac{(\tau-\beta_6)^{\beta_5}}{\beta_3}\right]  + \beta_4\right)\:\mathbbm{1}\{\tau > \beta_6 \}
\end{split}
\label{eq:acgshapemodel}
\end{align}
where $\tau \geq 0$, $\beta_2 + \beta_4$ is the peak height of the ACG, $\beta_6$ is the value of $\tau$ when $M(\tau)$ is maximized, $\beta_6$ is the baseline firing rate, i.e. the value of the ACG as $\tau \rightarrow +\infty$, $\beta_1$ is the slope rising out of zero lag, $\beta_3$ is the slope falling from the peak, and $\beta_5 \in \{1,2\}$ encodes the sharpness of the peak (characteristically, $\beta_5 = 1$ for pyramidal neurons and non-FS interneurons and $\beta_5 = 2$ for both FS and non-FS interneurons). 
Placing these in the vector $\vect{\beta}$ and denoting $A_{Data}(\tau)$ as the empirical ACG, we approximated the solution to,

\begin{equation*}
    \hat{\vect{\beta}} = \argmin_{\vect{\beta}} \sum_{\tau}(A_{Data}(\tau)-M(\tau,\vect{\beta}))^2
\end{equation*}
with Matlab's function \textit{lsqcurvefit} and a MultiStart object with 500 starting points for each ACG in the dataset. The MultiStart object attempts to find various local solutions in the cost function beginning with many random initializations.

 The LIF model parameters were fit to the medoids of the data clusters in Figure~{\ref{fig:biophysical-models}}. Medoids were determined by Matlab's \textit{k-medoids} function with 10,000 replicates. These random initializations once again search for many local minima and the best solution is chosen.
 To fit the \invivo ACGs, we tried to minimize the squared error between the simulated LIF neuron ACGs, $A_{LIF}(\tau,\vect{\gamma})$, and the sum of exponential ACG model (Eqs. 27). To make this problem reasonably tractable, we only fit the ACGs of the pyramidal and interneuron medoids, judging those to be representative examples. Denoting $rp$ as the fixed refractory period, let $\vect{\gamma}$ = ($\tau_T, V_i, \tau_I, \mu_I, \sigma_I, D_{\mu}, rp, (\mu_{max}-\mu_{min})/2, \alpha, \tau_m, V_t - V_i$). We explored various approximations of,
 
 \begin{equation*}
    \hat{\vect{\gamma}} = \argmin_{\vect{\gamma}} \sum_{\tau}(A_{LIF}(\tau,\vect{\gamma})-M(\tau,\hat{\vect{\beta}}))^2
\end{equation*}
  and reported $mean(\hat{\vect{\gamma}}) \pm std(\hat{\vect{\gamma}})$ across ten  independent runs of the differential evolution algorithm (docs.scipy.org) in the Results. The algorithm was run ten independent times for both the pyramidal fit and interneuron fit with a population size of 50 solution vectors and optimization parameters mutation = 0.7, recombination = 0.7, generations = 100. Hence, each optimization was comprised of (10 runs) X (50 solution vectors) X (100 generations) = (50,000 LIF simulations) per neuron type each having a trial duration of $1000^2$ ms. These optimization parameters were tuned by hand.
 
 To characterize \textit{CCG Sharpness} in Figure~{\ref{fig:monosynapse}}, we used simultaneous acceptance bands from a $\Delta$-jitter hypothesis test as a threshold to distinguish fine-timescale peaks from sampling variability derived from coarse-timescale dynamics, as described in Amarasingham et al. (2012)\nocite{Amarasingham2012}. Exact pointwise acceptance bands are derived by generating $M$ Monte Carlo surrogate CCGs from spike times that are interval-jittered in 
contiguous and disjoint intervals of duration $\Delta$. Exact simultaneous acceptance bands are more conservative and control the probability of {\it any} false rejections, across all lags (so-called `familywise' error rate [FWER] corrections). For all the panels in Figure~{\ref{fig:monosynapse}}, spike times were discretized into 0.1 ms elementary bins. Let $c_0(\tau)$ be the observed and unnormalized CCG at lag $\tau$ from two spike trains and let $c_1(\tau),c_2(\tau),...,c_M(\tau)$ be the surrogate CCGs derived from $M$ Monte Carlo surrogate spike trains. For each $\tau$, form the order statistics $\{c_{(j)}(\tau)\}$ by ordering (ranking) the statistics $\{c_j(\tau)\}$ such that $c_{(0)}(\tau) \leq c_{(1)}(\tau)\leq...\leq c_{(M)}(\tau)$. The pointwise acceptance region is simply the center $95\%$ of these values per $\tau$ (shown in Figure~{\ref{fig:monosynapse}} examples). To standardize across different lags, we compute

\begin{align*}
\begin{split}
v(\tau) &= \frac{1}{M-1}\sum_{m=1}^{M-1}c_{(m)}(\tau)\\
s(\tau) &= \sqrt{\frac{1}{M-2}\sum_{m}^{M-1}(c_{(m)}(\tau)-v(\tau))^2}\\
c_{m}^{*}(\tau) &= \frac{c_{(m)}(\tau)-v(\tau)}{s(\tau)}\\
\end{split}
\end{align*}
Now, let $c_{m}^{+} = \max_{\tau}c_{m}^{*}(\tau)$ and $c_{m}^{-} = \min_{\tau}c_{m}^{*}(\tau)$. Rank each again, such that, $c_{(0)}^{+} \leq ... \leq c_{(M)}^{+}$ and $c_{(0)}^{-} \leq ... \leq c_{(M)}^{-}$. The upper and lower simultaneous acceptance bands for an $\alpha$-level hypothesis test are,

\begin{align*}
\begin{split}
low^{*}(\tau) = &c_{(\frac{\alpha}{2}M)}^{-}s(\tau) + v(\tau) \\
up^{*}(\tau) = &c_{(M-\frac{M\alpha}{2})}^{+}s(\tau) + v(\tau).
\end{split}
\end{align*}
We used $\alpha=0.05$ in this study, and these control the FWER rate. Pointwise bands are appropriate if a single $\tau$ was viewed in isolation (typically for reasons of particular physiological significance), whereas the simultaneous bands control for 'fishing' for significance across lags (see Amarasingham et al., 2012, for more extensive discussion.) Finally, in Figure~{\ref{fig:monosynapse}}, we define,

\begin{equation*}
\textit{CCG Sharpness}(\Delta) = \sum_{\tau} \mathbbm{1}\{C_0(\tau)>up^{*}(\tau,\Delta)\}.
\end{equation*}
This expression is a function of $\Delta$ (elsewhere omitted for brevity). Both the observed CCGs and acceptance bands were scaled to units of Hz in the plots of Figure~{\ref{fig:monosynapse}}. We compare the \invivo CCGs with the constant threshold LIF and adaptive threshold LIF by considering the quantity $\textit{CCG Sharpness}(\Delta)$ for $\Delta \in \{0.1,0.2,...,90\}$ ms. We also found that using pattern jitter did not qualitatively change the conclusions of Figure~{\ref{fig:monosynapse} \citep{harrison2009rate,Amarasingham2012}. For these results (not reported here) the pattern jitter history parameter ($R$ in \citet{harrison2009rate}), which specifies the length of patterns that are preserved,  was adjusted individually for each pyramidal neuron, and set to a value such that the ACG fell within 1 Hz of $\beta_4$ (from Eq. \ref{eq:acgshapemodel}). This heuristic yielded a pattern jitter-corrected ACG that looks approximately flat, for the ACGs we examined.

%
\section*{Conflict of interest}

The authors declare that they have no conflict of interest.

\nolinenumbers

%
%
\bibliographystyle{spbasic}
\bibliography{monosynapse}

\section*{Supporting information}
\setcounter{equation}{0}

\subsection*{S1 Appendix: The injected synchrony model with short-term plasticity}

The simple short-term plasticity models of \citet{dayan2003theoretical} were used to study the accuracy of the injected synchrony model in the same manner as Figure~{\ref{fig:monosynapse-estimation}} in the main text. For a depressed synapse, the variable $P_{rel}$ encodes the synaptic efficacy, described by,
\begin{equation}
    \frac{dP_{rel}}{dt} = \frac{-P_{rel} - P0}{\tau_p} - f_D P_{rel} \sum_f \delta(t-t^{(f)})
\end{equation}

where the $t^f$ are the presynaptic spike times, $f_D$ is the strength of the depression, $\tau_p$ is the timescale, and $\delta(t) = +\infty$ when $t=0$ and $0$ otherwise. Similarly, a facilitated synapse is described by,
\begin{equation}
    \frac{dP_{rel}}{dt} = \frac{-P_{rel} - P0}{\tau_p} + f_F (1-P_{rel}) \sum_f \delta(t-t^{(f)})
\end{equation}
where $f_F$ is the strength of the facilitation.

Figure~{\ref{fig:plastic}} demonstrates the influence of the synaptic plasticity parameters on estimation accuracy. This bias is in general smaller than by the biophysical features studied in Figure~{\ref{fig:monosynapse-estimation}}. The implications of these results will be studied in future work.

\begin{figure*}
\includegraphics[width=.9\linewidth]{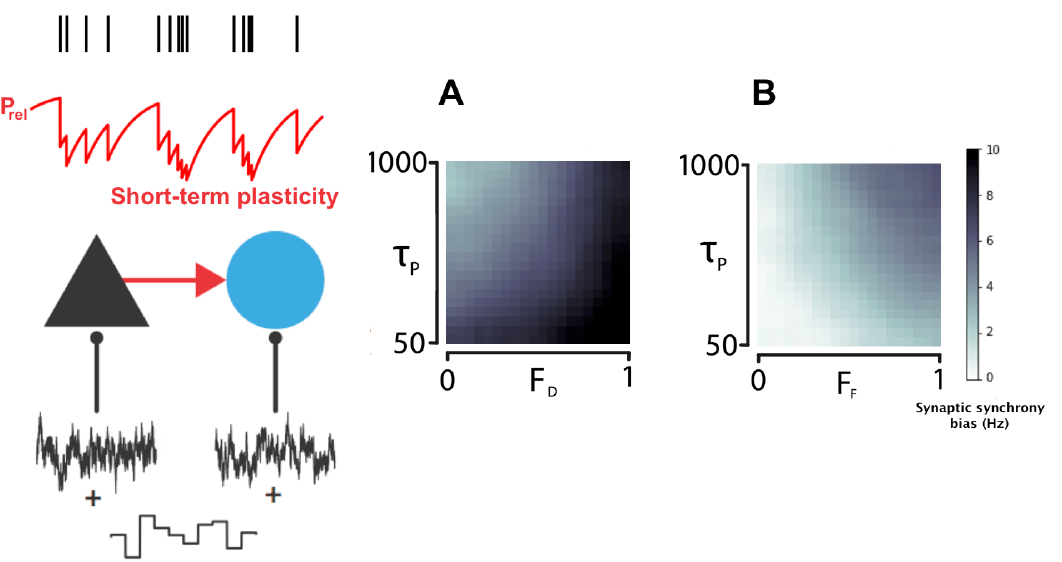} 
\caption{{\bf Synchrony estimation with short-term plasticity} 
The empirical bias of the synaptic synchrony rate estimator is studied as a function of short-term plasticity parameters. A: The timescale ($\tau_P$) and magnitude ($F_D$) of a depressed synapse. B: The timescale ($\tau_P$) and magnitude ($F_F$) of a facilitated synapse. Each pixel in each grid was estimated from a simulation of trial duration $10^8$ ms. Note that the colorbar is consistent, but different from Figure 5.}   
\label{fig:plastic}
\end{figure*}

\begin{figure}[!htbp]
\includegraphics[width=.9\linewidth]{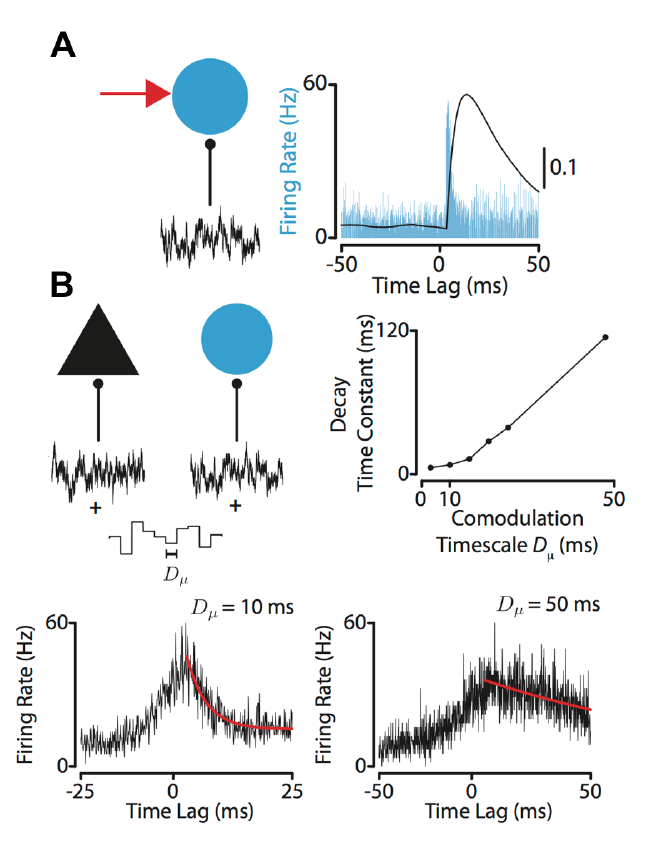} 
\caption{{\bf Mechanistic sources of CCG features.} 
{\bf A:} We considered here a single interneuron model, whose synaptic input conductance was triggered by
a single spike in each trial. The poststimulus time histogram of the spike count is shown in blue. The spike-triggered average of the postsynaptic membrane potential is the superimposed black line. Note the sharpness
of the poststimulus time histogram in comparison to the spike-triggered average. {\bf B:} (Top-Right) The decay time constant of the resulting CCG increases steeply with the comodulation timescale. (Bottom-Left) The decay time constant of the CCG is simply measured by fitting a single exponential decay to its positive lags part. For the value of $D_{\mu}$ used in this study and lower, the CCG appears symmetric. (Bottom-right) The CCG shape becomes asymmetric as $D_{\mu}$ gets larger, and decays more slowly, as observed in~\citet{Kobayashi2019}.}   
\label{fig:ccgmech}
\end{figure}

\subsection*{S2 Appendix: Additional dynamical properties of the biophysical model}

We describe a few mechanisms that may underlie other features of the CCG, as mentioned in the main text (Figure~{\ref{fig:ccgmech}}). We considered in Figure~{\ref{fig:ccgmech}A} a single interneuron model, whose synaptic input conductance was triggered by a single spike in each trial. The spike count was averaged over trials and normalized to produce the poststimulus time histogram. The postsynaptic membrane potential was measured relative to the
presynaptic spike time, and then averaged over trials (black line) to form the spike-triggered average. Note the sharpness
of the poststimulus time histogram, in comparison to the spike-triggered average. 

The overall shape of a CCG can be dependent on the timescale of firing comodulation. In Figure~{\ref{fig:ccgmech}B} we simulated  an interneuron and a pyramidal neuron receiving both shared and independent noise, with varying timescale values for the shared noise ($D_{\mu}$).

Finally, in Figure \ref{fig:m-p}, we further characterize dynamical features of the membrane potential (spike-triggered average and correlation functions).

\begin{figure}[!htbp]
\includegraphics[width=.9\linewidth]{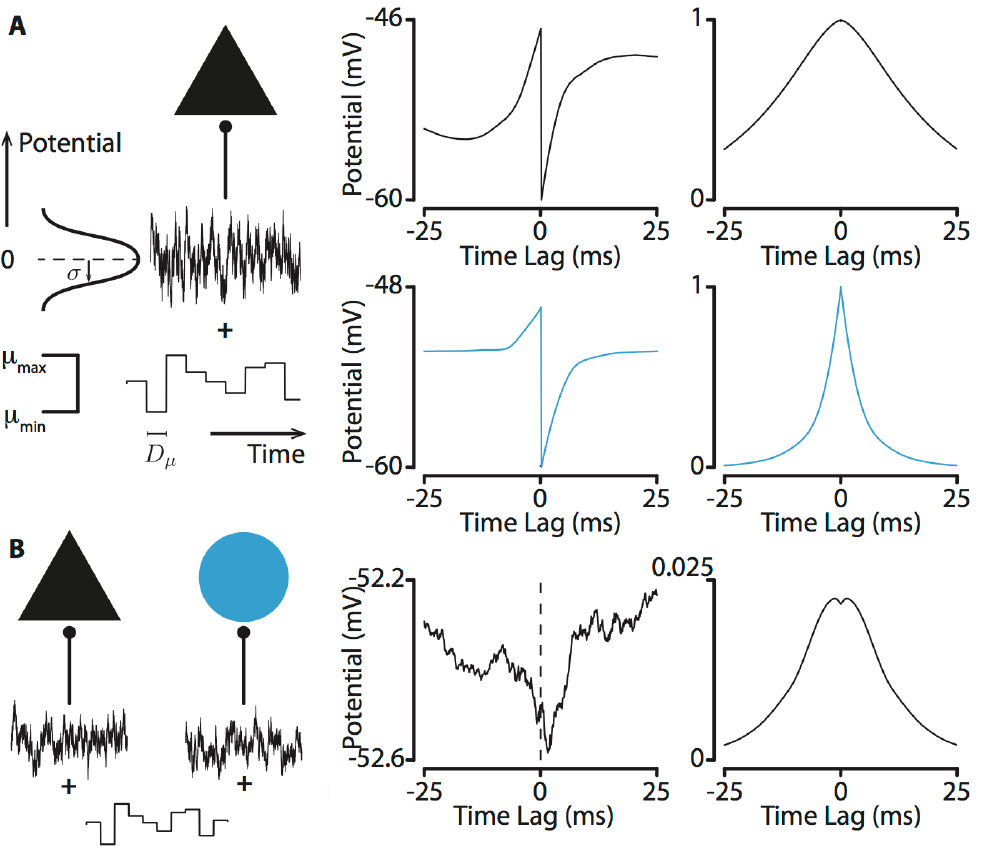}
\caption{{\bf Membrane potential dynamics for the biophysical models (Table 1 parameters). A:} The model of nonstationary firing based on a single neuron is depicted (pyramidal cell, top row; interneuron, bottom row). We show the spike triggered average of the membrane potential (left column) and
the auto-/cross-correlation function of the membrane potential (right column).
 {\bf B:} Model of co-modulated firing based on a
pyramidal cell (black triangle) and interneuron (blue disk) pair. Plots are as in A, with the pyramidal spikes
taken as reference.}   
\label{fig:m-p}
\end{figure}

\subsection*{S3 Appendix: Jitter and bootstrap sensitivity; some calculations}


As elsewhere, condition on the reference train, so we are jittering the target train. $\Delta$ is fixed. Denote by $M$ the number of intervals containing at least one spike in both trains. The goal is to  compare surrogate generation methods with respect to sensitivity of departures from the null detected by the synchrony statistic. Denote by $n_{t,i}$ the number of spikes for the target train in window $i$. Let $k_i$ be the number of bins in window $i$, which will contribute to the synchrony count upon containing a target spike (note this depends only on the reference train, and the definition of synchrony width). Then let us call $Y^{b/s}$ the synchrony count from a bootstrap surrogate, $Y^{b/s}_i$ the synchrony count from window $i$ in a bootstrap surrogate. $Y^{b/s}=\sum_{i=1}^M Y^{b/s}_i.$ $Y^j$ and $Y^j_i$ are, analogously, from a jitter surrogate. Let $\bf N$ be the interval counts. Then notice that
\begin{equation}
E[Y^{b/s}| {\bf N}] = E[Y^{j}| {\bf N}].
\end{equation}
Also,
\begin{align}
\sum_{i=1}^M \frac{n_{t,i}}{\Delta} \left( 1 - \frac{n_{t,i}}{\Delta} \right) k_i
&= Var[Y^{b/s}| {\bf N}]  \\
&>
Var[Y^{j}| {\bf N}] \\ &= \sum_{i=1}^M \frac{n_{t,i}}{\Delta} \left( 1 - \frac{n_{t,i}}{\Delta} \right) k_i \underbrace{\frac{\Delta - k_i}{\Delta - 1}}_{\text{Hyperg. c.f.}}
\end{align}
(The hypergeometric correction factor is the correction factor for sampling with, as opposed to without, replacement; analogous quantities give similar relationships in the continuum.)

Let us translate observations into standard deviations with respect to the common mean of the bootstrap and jitter distributions.  Suppose the observed synchrony is $\tilde{x}$. 

\begin{equation}
\frac{ 
\frac{ \tilde{x} - E[Y^{b/s}| {\bf N}] }{ \sqrt{Var[Y^{b/s}| {\bf N}] }} }
{ \frac{ \tilde{x} - E[Y^{j}| {\bf N}] }{ \sqrt{Var[Y^{j}| {\bf N}] }} } 
= \frac{\sqrt{Var[Y^{j}| {\bf N}]}}{\sqrt{Var[Y^{b/s}| {\bf N}]}}
= \frac
{ \sqrt{ \sum_{i=1}^M \frac{n_{t,i}}{\Delta} \left( 1 - \frac{n_{t,i}}{\Delta} \right) k_i \frac{\Delta - k_i}{\Delta - 1}}}
{ \sqrt{ \sum_{i=1}^M \frac{n_{t,i}}{\Delta} \left( 1 - \frac{n_{t,i}}{\Delta} \right) k_i}}
\end{equation}

It is instructive to consider constant $n_{t,i}=t$ and $k_i=k$ examples. Then 
\begin{equation}
\frac{Std[Y^{j}| {\bf N}]}{{Std[Y^{b/s}| {\bf N}]}} = \frac
{ \sqrt{ \sum_{i=1}^M \frac{n_{t,i}}{\Delta} \left( 1 - \frac{n_{t,i}}{\Delta} \right) k_i \frac{\Delta - k_i}{\Delta - 1}}}
{\sqrt{ \sum_{i=1}^M \frac{n_{t,i}}{\Delta} \left( 1 - \frac{n_{t,i}}{\Delta} \right) k_i }}
= \sqrt{ \frac{\Delta-k}{\Delta-1}},
\end{equation}
where $Std$ denotes standard deviation. That is, in the constant $t,k$ case, if $\tilde{x}$ is $m$ standard deviations from the mean with respect to the jitter distribution, it is $\sqrt{\frac{\Delta-k}{\Delta-1}}m$ standard deviations from the the mean with respect to the bootstrap distribution. This provides an explanation of the finer sensitivity of the jitter test, in the case of a synchrony test statistic. Note the key quantity involves $k/\Delta.$

\subsection*{S4 Appendix: A necessity argument for interval jitter tests of conditional uniformity}

Let $\bf X$ be the spike train(s) and $\bf N=\bf N(\bf X)$ be the interval counts.  Suppose that the null hypothesis is that $\bf X$ is conditionally uniform, given $\bf N$. Suppose one has a level $\alpha$ test of the null hypothesis specified by critical region $C$. Then it follows that
\begin{equation}
\Prob( C | \bf N=\bf n) \leq \alpha \quad {\rm for \, all } \,\, \bf n.
\end{equation}
[Because otherwise one can obtain a contradiction, as follows: suppose it is not true. Then there is a $\bf n'$ such that $\Prob( C | N=\bf n')) > \alpha.$ But then consider the distribution satisfying $\bf X$ uniform on the set $\{\bf N(\bf X)=\bf n'\}$. This distribution is both in the null and yields $\Prob(C) > \alpha$.]

This suggests that, taking the conditional  uniformity null for granted, any test is equivalent to a permutation test. We can see that the test determined by $C$ must be equivalent to a (deterministic) permutation test specified by a test statistic $T(\cdot)$. Define $C_k = C \cap \{\bf N=k\}$. Let
\begin{equation}
T({\bf X}) = \ind \{ {\bf X} \in C_{\bf N( X)} \}.
\end{equation}
It follows that
\begin{equation}
C_{T({\bf X}),\alpha} = \{ {\bf X} : T({\bf X}) \geq 1 \} = C,
\end{equation}
where $C_{T({\bf X}),\alpha}$ is the level-$\alpha$ deterministic permutation (i.e., interval jitter) test induced by statistic $T(\cdot)$. (The Monte Carlo version coincides with this deterministic test in the limit as the number of surrogates goes to $\infty.$)

\begin{figure}[!htbp]
\includegraphics[width=.9\linewidth]{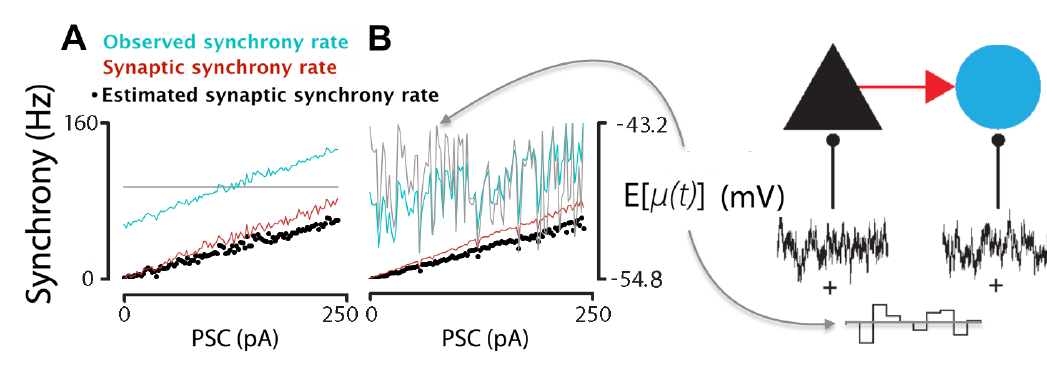} 
\caption{{\bf Recovering monotonicity of the synaptic synchrony with respect to the PSC. A:} We considered 100 equally spaced postsynaptic current amplitudes (PSC $\in [0,240]$ pA). In this first panel the nonstationary mean input itself has mean, $E[\mu(t)] = -48$ mV (from Table 2 parameters). Notice in this case, the monotonicity of the observed synchrony rate ($S/(|\vect R| \cdot 10^{-3}$) is already deconfounded from the network state. The estimated synaptic synchrony rate ($\hat{\theta}/(|\vect R| \cdot 10^{-3})$) approximates the synaptic synchrony rate ($\theta/(|\vect R| \cdot 10^{-3})$) (the bias for this estimate was systematically explored as a function of various biophysical parameters in Figure~{\ref{fig:monosynapse-estimation}}). {\bf B:} For each PSC, we simulated one trial ($1,000$ s duration). For each such trial, $\mu_{\rm min}$ and $\mu_{\rm max}$ were randomly sampled such that $(\mu_{\rm min}+\mu_{\rm max})/2=E[\mu(t)]$ was distributed uniformly between -53 and -43 mV [$E[\mu(t)] \sim U(-53,-43)$ mV]. Per trial, this induces a non-monotonic relationship between the observed synchrony rate and the PSC. The observed synchrony rate is in fact highly correlated with $E[\mu(t)]$ across trials (r =0.88), but not with the PSC (r=-0.12). However, the injected synchrony model deconfounds the monotonically increasing PSC, yielding high correlation between the estimated synaptic synchrony rate and the PSC (r = 0.990).}   
\label{fig:conf}
\end{figure}
\subsection*{S5 Appendix: Monotonicity of synaptic synchrony with respect to the postsynaptic current amplitude}

In section \ref{sec:biophysics-monosynapse}, we pose monosynaptic inference as the task of correctly identifying the monosynapse's causal contribution to the CCG. The significance of this causal estimation is that it implies information about the monosynapse in the spike train has been deconfounded from the background network in a robust way, and this might be of broad use for \invivo experiments. 

To illustrate this point in a more concrete setting, Figure~{\ref{fig:conf}} shows that a strong linear correlation can be recovered between $\hat{\theta}$ and the postsynaptic current amplitude (PSC) even when the correlation is explicitly confounded by background network activity. In Figure~{\ref{fig:conf}A} the observed synchrony rate (derived from the uncorrected CCG) is already monotonic with the PSC (because the background parameters are equal across each stratum of the PSC). However, Figure~{\ref{fig:conf}B} demonstrates that, using $\hat{\theta},$ nearly the same result obtains when the background parameters are randomized for each level of the PSC. 

\end{document}